\documentclass[%
 reprint,
superscriptaddress,
 amsmath,amssymb,
 aps,
]{revtex4-2}
\usepackage{xcolor}
\usepackage{graphicx}
\usepackage{dcolumn}
\usepackage{bm}
\usepackage{gensymb}
\usepackage{breqn}
\usepackage{lineno}

\begin{document}
\preprint{}

\title[Computational Ultrasound Imaging]{Four-Dimensional Computational Ultrasound Imaging of Brain Haemodynamics}

\author{Michael D. Brown}
 \email{m.brown@erasmusmc.nl}
 \affiliation{$Department\ of\ Neuroscience,\ CUBE,\ Erasmus\ MC,\ Rotterdam,\ Netherlands$}
 \affiliation{$Department\ of\ Medical\ Physics\ and\ Biomedical\ Engineering,\ University\ College\ London,\ London,\ United\ Kingdom$}

\author{Bastian S. Generowicz}
 \affiliation{$Department\ of\ Neuroscience,\ CUBE,\ Erasmus\ MC,\ Rotterdam,\ Netherlands$}

\author{Stephanie Dijkhuizen}
 \affiliation{$Department\ of\ Neuroscience,\ Erasmus\ MC,\ Rotterdam,\ Netherlands$}

\author{Sebastiaan K.E. Koekkoek}
 \affiliation{$Department\ of\ Neuroscience,\ Erasmus\ MC,\ Rotterdam,\ Netherlands$}

 \author{Christos Strydis}
 \affiliation{$Department\ of\ Neuroscience,\ CUBE,\ Erasmus\ MC,\ Rotterdam,\ Netherlands$}
 \affiliation{$Department\ of\ Quantum\ and\ Computer\ Engineering,\ TU\ Delft,\ Netherlands$}%
 \author{Johannes G. Bosch}
 \affiliation{$Department\ of\ Cardiology,\ Thorax\ Biomedical\ Engineering,\ Erasmus\ MC,\ Rotterdam,\ etherlands$}

  \author{Petros Arvanitis}
 \affiliation{$Department\ of\ Neuroscience,\ CUBE,\ Erasmus\ MC,\ Rotterdam,\ Netherlands$}

 \author{Geert Springeling}
 \affiliation{$Experimental\ Medical\ Instrumentation,\ Erasmus\ MC,\ Rotterdam,\ Netherlands$}

 \author{Geert J.T. Leus}
 \affiliation{$Signal\ Processings\ and\ Systems, Department\ of\ Microelectronics,\ TU\ Delft,\ Delft,\ Netherlands$}

 \author{Chris I. De Zeeuw}
 \affiliation{$Department\ of\ Neuroscience,\ Erasmus\ MC,\ Rotterdam,\ Netherlands$}
 \affiliation{$Netherlands\ Institute\ for\ Neuroscience,\ Royal\ Dutch\ Academy\ for\ Arts\ and\ Sciences,\ Amsterdam,\ Netherlands$}

\author{Pieter Kruizinga}
\affiliation{$Department\ of\ Neuroscience,\ CUBE,\ Erasmus MC Rotterdam,\ Netherlands$}

\date{\today}

\begin{abstract}
Four-dimensional ultrasound imaging of complex biological systems such as the brain is technically challenging because of the spatiotemporal sampling requirements. We present computational ultrasound imaging (cUSi), a new imaging method that uses complex ultrasound fields that can be generated with simple hardware and a physical wave prediction model to alleviate the sampling constraints. cUSi allows for high-resolution four-dimensional imaging of brain haemodynamics in awake and anesthetized mice. \\
\end{abstract}

\maketitle

\section{Introduction}
The advent of ultrafast ultrasound ($>$5000 frames per second (fps)) imaging over the last 20 years, enabled by increased computational power and parallel receive electronics, has spurred the development of multiple novel imaging modes for biomedical ultrasound~\cite{Tanter2014,Couade2016}. The formation of complete images within a short ($<$1 ms) temporal window allows for accurate quantification of tissue, blood, and contrast agent motion. This facilitates measurement of tissue elasticity and arterial stiffness~\cite{Sebag2010,Vappou2010}, super-resolution via localisation and tracking of individual micro-bubbles~\cite{Errico2015,Christensen2020}, and vastly enhanced imaging of blood flow over a wide field-of-view~\cite{Bercoff2011}. The latter has resulted in the emergence of functional ultrasound imaging (fUS or fUSi) a neuro-imaging technique  - which is able to detect small changes in cerebral blood volume (CBV) induced by neurovascular coupling~\cite{Mace2011,Deffieux2021}. In comparison to other neuro-imaging modalities such as fMRI, fUS offers greater ease of use at significantly lower cost while delivering a higher spatial-temporal resolution, with a recent demonstration, in conjunction with contrast agents, of its capability to detect vascular activity with a 6.5 $\mu m$ spatial resolution~\cite{Renaudin2022}. 

Ultrafast ultrasound imaging, however, remains, principally, a two-dimensional technique due to the stringent spatiotemporal sampling requirements of the imaging process. This imaging process necessitates the transmission of a sequence of planar or diverging waves at high frame rates ($\geq$5kHz) while recording, in-parallel, the back-scattered signals over a surface sampled spatially and temporally at Nyquist rates~\cite{Tanter2014}. In the case of 3D imaging, this typically requires thousands of elements (as compared with 64-256 for 2D imaging) and a corresponding number of independent data channels with associated radio frequency (RF) digitisers. Recent works have reported 1024 channel systems for 3/4D cardiac imaging~\cite{provost20143d,petrusca2017new}, super-resolution~\cite{chavignon20213d,heiles2019ultrafast}, and functional imaging in rats~\cite{Rabut2019}. However, these required the use and synchronisation of multiple data acquisition systems which is both extremely costly and technically complex, making it infeasible for most clinical applications. To reduce the number of necessary independent data-channels two approaches are conventionally considered: 1) retaining a fully-sampled array but using more complex readout schemes employing application-specific integrated circuits (ASICs) to combine and pre-beamform signals~\cite{Chen2017,janjic20182,sauvage20194d}, 2) sparse arrays that strategically sub-sample an aperture with an element distribution designed to minimise side-lobes~\cite{wei2021high,Ramalli2022}. However, both methods impose compromises on the image formation process and such arrays are technically complex to realise. Moreover, the small element sizes (on the order of the acoustic wavelength $\lambda \sim100-300 \mu m$) imposed by Nyquist on all conventional arrays result in poor individual element sensitivity. For haemodynamic imaging, particularly in small animals that suffers from small signal amplitudes, this poses a significant challenge. 

\begin{figure*}
    \centering
    \includegraphics[width=1.9\columnwidth]{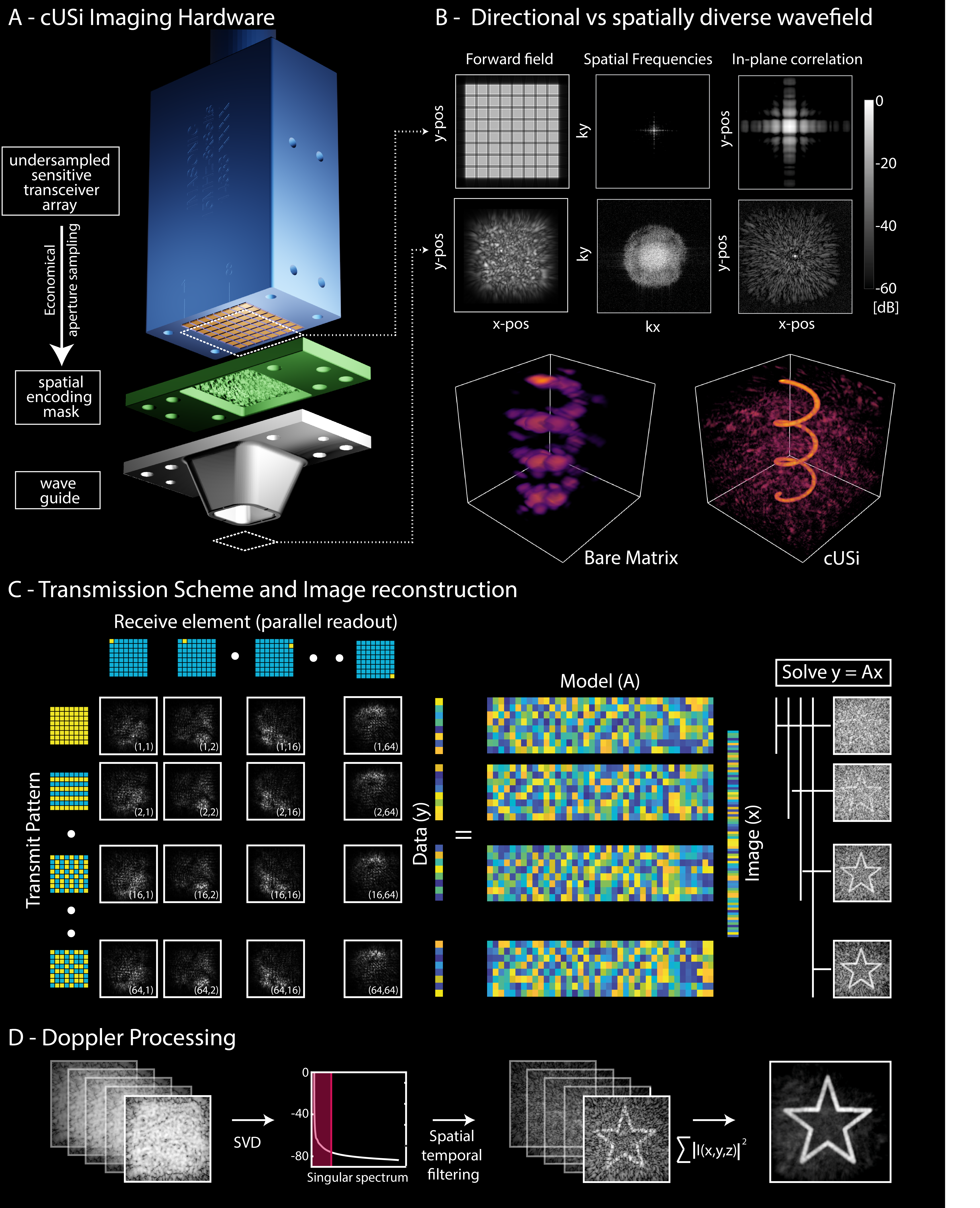}
    \caption{(A) Rendering of the cUSi imaging hardware. A matrix probe populated with acoustically large, sensitive, elements is modified by attaching a smart, plastic, encoding mask that scrambles the transmit and receive wave-fields. The mask is coupled to a waveguide that confines the transmitted fields and provides the necessary imaging offset. (B) Scrambling the transmitted wave-field provides a broader sampling of k-space and allows us to trade lateral resolution and side-lobe intensity. (C) The probe is driven using a Hadamard-encoded synthetic aperture scheme, signals from different transmissions are separately reconstructed then coherently summed to form each 3D volume. Images are reconstructed using a model-based approach. The system response is calibrated with a one-time measurement and then correlated with each set of measurements to recover an image. (D) To generate Doppler images, we continuously transmit to acquire data that is used to reconstruct separate volumes at a rate of $\sim$400 Hz. The data is spatiotemporally filtered and the power associated with blood flow in each voxel over time is evaluated to form the PDI.}
    \label{Figure_1}
\end{figure*}

To address the aforementioned challenges, we introduce computational ultrasound imaging (cUSi). The complex hardware requirements and sensitivity challenges associated with sparse or ASIC addressed arrays are avoided with the use of a simple, fully populated, matrix probe that is under-sampled using acoustically large, and therefore highly sensitive, elements. Inherently these large elements are extremely directional (Fig. 1A) and thus incapable of high-resolution imaging. To compensate for this loss of resolution, we attach a smart, plastic encoding mask to the probe which scrambles the transmitted and received wave-fields of every element. By modifying the field with a random encoding mask, we more evenly sample the k-space of our imaging aperture while avoiding any symmetries that would give rise to artefacts such as grating lobes (Fig. 1B) ~\cite{favre2023transcranial}. This more uniform sampling of k-space increases the lateral resolution of the imaging system at the expense of increased clutter/side-lobes. The challenge introduced, however, by imaging with complex wave-fields is that it prohibits the use of the conventional, geometry-based, processing (e.g., delay-and-sum) that is routinely used in ultrasound for image formation~\cite{Szabo2004}. Instead, we reconstruct images using a model-based approach after calibrating the 3D imaging response of our system using a one-time measurement (Fig 1C and Materials and Methods) ~\cite{Kruizinga2017}. 

cUSi falls into the category of computational imaging which covers techniques across a range of modalities that broadly aim to realise cheaper and/or faster imaging devices in part by shifting the burden of image formation from complex hardware onto computation~\cite{Antipa2018,Malone2023,Tondo2017,Fromenteze2015,edgar2019principles,mait2018computational}. Within the ultrasound domain the exploitation of reverberant media to reduce the number of sensors required for imaging has been investigated since the 90's~\cite{draeger1999one2,derode1999ultrasonic,Montaldo2004,Robin2018}. However, the translation of these methods to in-vivo imaging has, traditionally, been hindered by challenges in separating back-scattered signals from the transducer cross-talk as well as the high-sensitivity of these media to small perturbations (e.g., temperature shifts). Recent work has demonstrated the application of reverberant media for 2D in-vivo Doppler imaging, however, as with earlier works ~\cite{Montaldo2004}, this utilised separate elements for transmit and receive and required the use of contrast agents due to poor SNR ~\cite{caron2023ergodic}. In photoacoustics, where back-scattered cross-talk is not present, reverberant media have been successfully applied to ultrafast imaging of haemodynamics and functional activity via changes in optical absorption~\cite{Li2020,Li20202,Li2021}. This current work builds on proof-of-principle work on the use of a spatial encoding mask to mitigate sampling constraints  ~\cite{Kruizinga2017,janjic2018structured}. Here, we show for the first time a practical implementation that translates to 3D in-vivo recordings and ultrafast imaging. 
 \section{Results}
We validated the potential of cUSi for in-vivo imaging of brain haemodynamics in both awake and anesthetised mice. In both experiments imaging was performed through a cranial window, which was covered with a TPX film (CS Hyde Company, IL, USA) in the awake case only. This craniotomy was applied to remove the distorting and attenuating effects of the skull on the wave-field. An 8x8-element matrix probe (1.25 mm pitch, 1x1 cm aperture, 13.8 MHz, Imasonic France) was used for all experiments. This was under-sampled by a factor of almost 500 compared to a fully populated array sampled at Nyquist rate of 32,000 sensors. The spatial encoding mask was fabricated in-house from Rexolite 1422 (owing to its favourable acoustic properties) via CNC micromachining. A aluminium waveguide was used to provide the required imaging offset and confine the transmitted field to the cranial window. The probe was driven using a synthetic-aperture transmission scheme while applying Hadamard encoding to boost SNR and attain full-information from the element array~\cite{Jensen2006,Chiao1997}. Each transmit-receive acquisition decreased the spatial correlations between voxels in the imaging domain improving reconstruction quality (Fig.\ 1C and Supplementary Fig.\ \ref{SFig_1}). The transmission rate was 32 kHz which, including a short dead-time for data-transfer and storage, resulted in a volume rate of 407 Hz. We acquired data continuously at this volume rate for 60 seconds generating a data ensemble from which, after removing frames subject to instability, the Doppler images were formed.

Prior to image reconstruction we applied a spatiotemporal filter to the data ensemble to obtain the signals arising from blood-flow, eliminating the large component originating from static soft tissue (Fig. 1D) ~\cite{demene2015spatiotemporal}. For the anesthetised mouse brain, volumes of size 6.8$\times$9.2$\times$8 mm with an isotropic voxel size of 40 $\mu m$ were reconstructed by correlating the measurements with a calibrated model of the spatiotemporal impulse responses of our imaging system (Materials and Methods) providing a 4D volume ensemble data-set $u(x,y,z,t)$. To form a power-Doppler image (PDI) we computed the average power for each voxel. To evaluate the direction of flow we computed a lag-1 auto-correlation of the temporal course of each voxel~\cite{madiena2018color}.  

In Fig.\ 2, we highlight the capability of cUSi for capturing brain haemodynamics in a small rodent model. In Fig.\ 2A, a 3D rendering of the entire PDI is shown: where both small and large vessels are resolved throughout the entire cranial window. In Fig.\ 2B, we feature the information on flow direction that can be extracted from the Doppler ensemble showing draining cortical vessels. Fig. 2C shows axial, coronal, and sagittal slices through the power Doppler volume indicating the richness of the small vessels that can be detected, particularly in the cortex, as well as the isotropic lateral resolution that we achieve. Finally, Fig.\ 2D demonstrates the dependency of the imaging performance on increasing Doppler ensemble size. In the current implementation, we require a 1-2 second acquisition time to form reliable volumes that could be used for other modalities such as fUSi. Lateral, Sagittal and Coronal flythroughs of the converged PDI are shown in Supplementary Mov.\ 1-2. In Supplementary Fig.\ \ref{SFig_2}, the power and colour Doppler volumes of an awake mouse are shown. Similar performance is obtained, however, there is a loss of fine detail in the cortex and of deeper vessels due to the loss in SNR. This loss of SNR is the result of attenuation generated in a TPX film that was used to cover the imaging window.
\begin{figure*}
    \centering
_    \includegraphics[width=2\columnwidth]{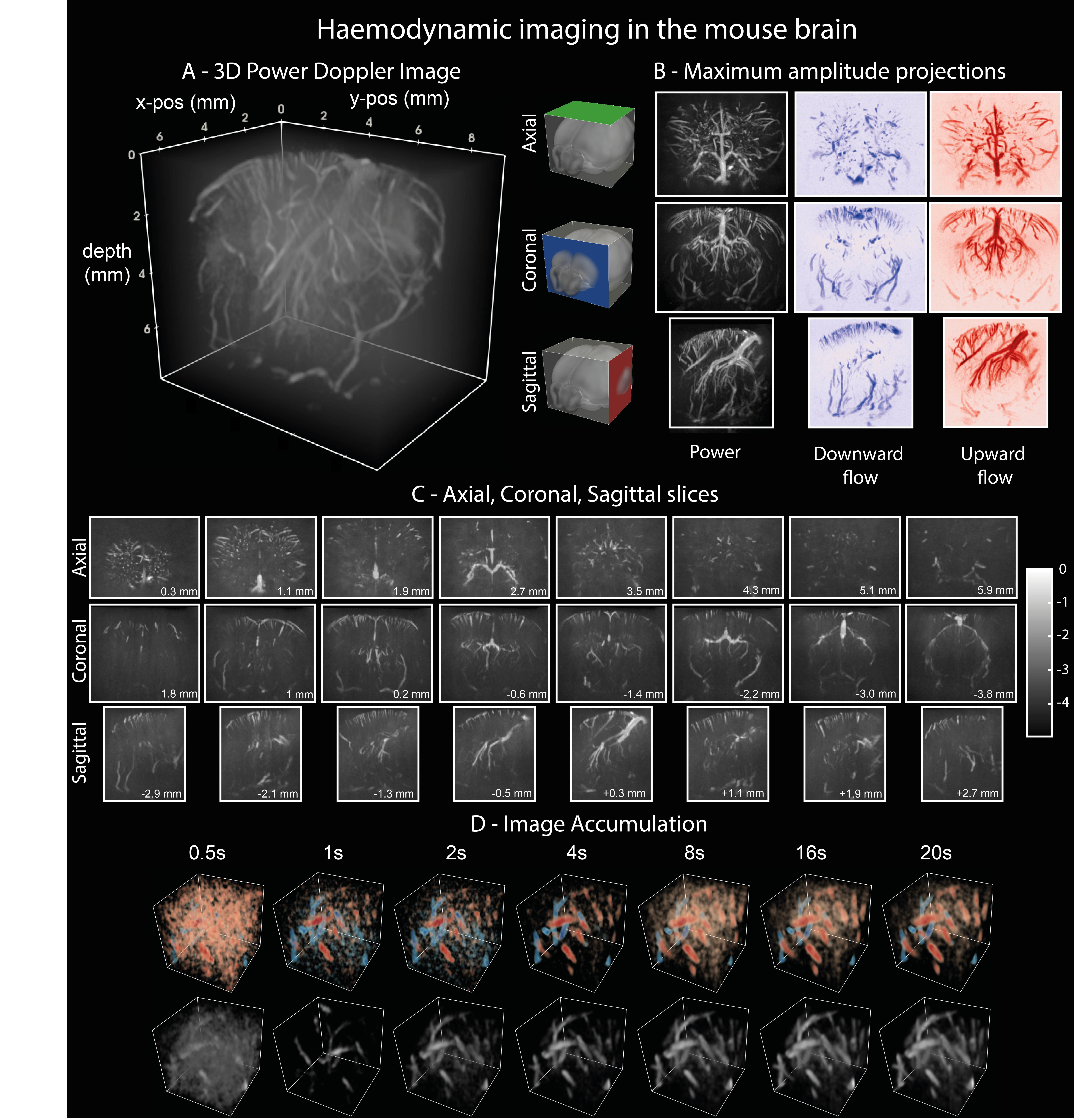}
    \caption{Results of Haemodynamic imaging in the anesthetised mouse brain. (A) 3D rendering of the reconstructed PDI of the anesthetised mouse brain. Image was formed by compounding a filtered data-set comprised of 8041 volumes. (B) Axial, coronal, and sagittal maximum amplitude projections through the reconstructed Power (left) and Colour (right) Doppler volumes. (C) Sub-projections through the PDI rendered in (A). Each slice was formed from a maximum amplitude projection through a set of planes with a thickness of 480 $\mu m$ (corresponding to 12 planes in the reconstructed volume). The approximate position of each projection in the brain is denoted on the image. For the coronal slices the positions are indicated relative to Bregma. (D) 3D rendering of vessels (bottom) and flow-direction (top) in a fixed region of the cortex formed via summation of an increasing number of frames. The time indicated assumes an underlying volume rate of 400 Hz (lower than that achieved in the experimental measurements) but neglects any frames removed to eliminate motion artefacts.}
    \label{Figure_2}
\end{figure*}

\section{Discussion}
Traditionally, the clinical paradigm for ultrasound imaging has focused on robust hardware and simple, fast, processing that provides real-time feedback. However, there are clear economical and technical challenges with scaling this approach to 3/4D imaging that limit its use to specific targets (e.g., cardiac imaging). Moreover, there are emerging clinical applications, such as low-cost wearable devices~\cite{hu2023wearable} and trans-cranial imaging~\cite{demene2021transcranial} for which this traditional paradigm is less appropriate. In this work, we have demonstrated for the first time the possibility for high-resolution 4D imaging of haemodynamics in the mouse brain using cUSi. This represents a proof-of-concept for such model-based approaches in biomedical ultrasound. In the future by designing custom hardware incorporating specific knowledge of the image reconstruction and clinical target there is the potential to realise computational imaging devices that address these contemporary clinical challenges.

The limitation in this work was the high sidelobes that originated from the large factor by which the array was under-sampled. This compressed the dynamic range compared to conventional 2D power Doppler. We were able to image the vasculature throughout the mouse brain with high resolution despite this. In the future, this dynamic range could be improved by firstly scaling up the number of matrix elements to bring it in-line with existing 2D systems, by further tuning of the physical media used to modulate the transmit/receive fields for the ultrasound array, and through joint optimisation of the driving scheme, mask design, and image reconstruction, particularly, leveraging developments in inverse design~\cite{molesky2018inverse,kellman2019physics} and deep-learning based image reconstruction~\cite{ongie2020deep}. 

\begin{acknowledgments}
This work was supported by the NWO-Groot grant CUBE (Grant no. 108845), and by the TTW-OTP grant TOUCAN (Grant no. 17208) financed by the Dutch Research Council (NWO).
\end{acknowledgments}
\cleardoublepage

\renewcommand{\thefigure}{S\arabic{figure}}
\setcounter{figure}{0}

\section{Materials and Methods}
\subsection{Animal Preparation.}
In this study we assessed cUSi in-vivo imaging experiments in both anesthetized (n=1) and awake (n=1) adult C57BL6/J mice (8-10 weeks old). At arrival animals were group housed under a 12/12h light/dark cycle, with controlled temperature and humidity, and with access to water and food ad libitum. After surgery, mice were individually housed. The national authority (Centrale Commissie Dierproeven, The Hague, The Netherlands; license: AVD1010020197846) granted ethical approval prior the experiments, which were performed according to the institutional, national, and European Union guidelines.

Surgery:
For both experiments a craniotomy was performed to facilitate imaging without distorting and attenuating effects of the skull. During surgery and the non-awake imaging experiments, mice were anesthetized using a isoflurane/oxygen mixture (5\% induction, 1.75-2\% maintenance), while body temperature was kept constant at 37 \textdegree C, and heart and respiration rate were monitored (Small Animal Physiological Monitoring System, Harvard Apparatus, MA, USA). The same device was used to fixate and level the head while drilling (Foredom) the cranial window. For the anesthetized imaging experiments a cranial window of 1x1 cm was made. For the awake experiments a smaller cranial window (+2 mm by -4 mm from Bregma, and ±4 mm width) was performed due to the placement of a pedestal (1x0.8 cm), which ensured head fixation in the experimental set-up during imaging. In the awake case only, the cranial window was covered with a TPX film (CS Hyde Company, IL, USA) and post-operative mice received 3-5 days of antibiotics (Baytril, 25mg/ml, Bayer, Germany) to prevent inflammation of the brain. Prior to the start of the imaging experiments, both the exposed and covered brain tissue was sprinkled with saline solution, after which ultrasound transmission gel (Aquasonic 100, Parker Laboratories, NJ, USA) was applied for acoustic contact between the brain and the the 8x8-element matrix probe positioned straight above the cranial window. 

\subsection{Matrix Probe.} Experiments used a custom built 64-element matrix probe (Fig.\ \ref{SFig_3}) (8$\times$8 square elements, 1.25 mm lateral width, 10$\times$10 mm aperture, 13.8 MHz central frequency, 65$\%$ -6 dB bandwidth, Imasonic, France). The element size was approximately 132$\lambda^{2}$ significantly larger than both the 16$\lambda^{2}$ used for 1D linear arrays (which while sampled at closed to Nyquist rate in-plane have a large elevational size) and the 4$\lambda^{2}$ used previously for 3D functional imaging of rats~\cite{Rabut2019}. The front-surface of the probe had a 15$\times$25 mm footprint and included tapped holes on each corner for fixation of coding masks to the front surface. 

\subsection{Spatial Encoding Mask.} For cUSI we aim to perturb the field for each matrix element by introducing a spatially varying phase shift analogous to an (optical) spatial light modulator or acoustic hologram~\cite{Melde2016}. We achieve this using a plastic coding mask that introduces a local, thickness-dependent delay to the transmitted field. 

The coding mask material, ideally, would fit a number of criteria. Namely, impedance matching to water and low acoustic attenuation to minimise insertion losses, high sound speed contrast to soft-tissue to maximise phase delays, and mechanically machineable/castable/printable with high precision. Rexolite 1422 was selected to balance these competing requirements as it possesses low acoustic attenuation and good acoustic matching to water (soft-tissue) while being sufficiently rigid to machine~\cite{Lopes2017}. 

The mask had a thin, smoothly varying, profile inspired by optical diffusers previously applied for 3D optical imaging~\cite{Antipa2018}. We generated a smoothly varying surface profile with an average lateral feature size of approximately 370 $\mu$m and a total height variation of 1.2 mm. The lateral feature size was constrained by the manufacturing method and the tool size while the ideal height scaling was determined empirically. We found that for masks that were too thin we insufficiently modulated the field of each element resulting in a broad point spread function (PSF) that was unable to resolve micro-vasculature. For masks that were too thick the increased losses due to greater internal reflection and increased absorption resulted in an SNR that was too low for imaging. 

The encoding mask design challenge is similar to that of sparse array design in ultrasound where considerable effort is often devoted to optimising the element configuration prior to fabrication~\cite{Ramalli2022}. With cUSi, however, we can retrospectively (and cheaply) modify our transmit/receive fields, and therefore the side-lobe distribution, simply by re-designing the encoding mask. While in this work we empirically designed our encoding mask, in the future, inverse-design principles~\cite{molesky2018inverse} could be adopted to refine this approach by incorporating the system physics and reconstruction nonlinearities into the encoding mask optimisation~\cite{markley2021physics,kellman2019physics}.

The encoding mask was generated for a 12$\times$12 mm area including a 1 mm lateral buffer around the active surface to eliminate sharp discontinuities that would create diffraction effects. This profile was exported as a point cloud to  SolidWorks (SolidWorks 2018, Dassault Systèmes, Vélizy-Villacoublay, France) where it was converted to a solid part that was interpretable to the CNC software. A 0.4 mm axial offset was added to the mask profile for mechanical stability and its aperture was expanded to match the waveguide opening. To attach the mask to the probe a 2 mm thick 15$\times$35 mm buffer was added around the encoding mask including four clearance holes to fixate it to the probe. The mask was then fabricated using a computer numerical control (CNC) machine (P60 HSC, Fehlmann) using a 200 $\mu m$ drill-bit. A photograph of the coding mask can be seen in Fig. \ref{SFig_3}.

\subsection{Waveguide.}
A tapered waveguide was added to the coding mask to further modify the transmitted wave-field. This was introduced for two reasons. First, the waveguide allows for a controlled offset between the probe surface and the imaging medium. From analysis of the system matrix the PSF was found to degrade significantly close to ($<$6 $mm$) the probe (Supplementary Fig. \ref{SFig_4}A). This occurs as the field of each matrix element needs to propagate a certain distance after the encoding mask in order to diverge such that it overlaps with that of neighbouring elements. Close to the probe each voxel is seen by as few as 4 elements which results in an extremely poorly conditioned reconstruction. To eliminate this an offset between the probe and imaging target is required with a distance determined by the divergence to the field introduced by the encoding mask and the element size. For this work, we found 12 mm to be sufficient. Second, the cranial window that can be safely introduced to the mouse's skull is limited, sagitally and laterally, to $\sim$8$\times$6 mm which is smaller than the probe aperture (10$\times$10 mm). As such a large fraction of the transmitted energy ($\geq40\%$) falls outside the imaging window. This is compounded by the additional divergence introduced by the encoding mask. With the addition of a waveguide this energy can be funnelled onto the desired imaging window (Supplementary Fig.\ \ref{SFig_4}B). 

The waveguide had input and output apertures of 14$\times$14 mm and 7$\times$7 mm respectively and a length of 10 mm. The minimum length was set by the discussed need for an offset for the imaging target as well as the requirement to avoid a sharp taper angle that would trap waves within the waveguide and reflect them toward the probe surface. There is no physical constraint on maximum length, however, an overly long waveguide suffers greater losses from absorption and, practically, is more prone to trapped air pockets in the ultrasound gel, which modify the system response from the calibration. The waveguide was fabricated from Aluminium using CNC micro-machining. Aluminium has a significantly higher acoustic impedance than both water and ultrasound gel (18 vs 1.5 MRayls) so near fully confines the transmitted field (Supplementary Fig.\ \ref{SFig_4}B).

\subsection{Transmission Scheme.}
A synthetic aperture transmission scheme~\cite{Jensen2006} (also referred to as full-matrix capture~\cite{Holmes2005}) was adopted for imaging. Each element sequentially transmits whilst recording the back-scattered signals on all the array elements in parallel allowing for each voxel in the imaging medium to be synthetically focused on during reconstruction. For the probe in this work each volume is therefore formed with a set of 64 transmissions. The principal disadvantage of synthetic aperture is poor SNR as each transmission only uses a single element. To improve this we applied spatial coding on transmit~\cite{Chiao1997}. For each of the 64 transmissions a different spatial code was applied to the matrix elements with each forming a column of a 64x64 matrix invertable matrix. We chose a Hadamard matrix as it is comprised of +/-1's so is easily implemented in hardware by flipping the polarity of the driving signal. By transmitting on all 64 elements each time we gain a factor of $\sqrt{N}$ or $\times$8 improvement in SNR. 

For conventional ultrasound it is necessary to decode the signals prior to reconstruction by applying the inverse of the Hadamard matrix. However, our image reconstruction fully models the relevant wave-physics so this step is not required. Instead we incorporate the spatial codes into our model of the transmission. cUSi would be equally feasible with alternative transmission schemes, for example, sequential focusing on each point in the imaging medium via time-reversal ~\cite{Robin2018}. While this would result in greater SNR it would not be compatible with the volume rate necessary for ultrafast imaging of blood flow. 

\subsection{Experimental Setup} 
The probe, encoding mask, and waveguide were each submerged and then assembled underwater using the four tapped holes on the front surface of the probe (Fig. 1A). Prior to assembly, if small air pockets were present on the mask, they were manually were manually removed from the encoding mask surface under a microscope using a syringe. Ultrasound coupling gel was then syringed into the end of the waveguide while still submerged to allow it to retain water when vertically mounted. Prior to application this syringe containing gel was centrifuged at 6000 rpm for 5 minutes to remove bubbles~\cite{bertolo2021whole}. The probe was then suspended vertically above the cranial window and lowered onto the exposed mouse-brain using a manual translation stage. Prior to imaging we assessed whether any air pockets were present in the waveguide by transmitting an impulse from each element and recording the back-scattered signals. The back-scattered signals were analysed to confirm that no signals originated within 10 $mm$ of the probe surface. 

\subsection{Experimental Measurements.} 
The probe was driven using an ultrasound research system (Vantage 64-LE, Verasonics Redmond, WA, USA) with a 5 cycle tone-burst centred on 15.625 MHz. A single transmit sequence was used for both the awake and anesthetised mice imaging experiments. The pulse-repetition frequency (PRF) was 32 kHz giving an underlying volume rate of 500 Hz. Transmissions were repeated in blocks of 6400 firings corresponding to 100 volumes. Between each block a dead time of approximately 50 ms was used to transfer the raw data buffer and write to a hard disk. The scheme was repeated for 60 seconds for both experiments resulting in 24400 volumes and an effective volume rate of 407 Hz. 

\subsection{Data pre-processing.}
Signals were sampled in 50$\%$ bandwidth (BW) mode to reduce data requirements. For each transmission 512 samples at 50$\%$ bandwidth~\cite{kaczkowski2016bandwidth} corresponding to an imaging depth of 24.6 mm were recorded on each element. The 50$\%$ BW data was then Fourier transformed and downsampled by a factor of 2 to generate a set of 128 frequencies $N_{\omega}$ that were used for reconstruction. These were evenly spaced between 11.7-19.4 MHz (i.e., a 50$\%$ bandwidth centred on 15.625 MHz). 

Next any frames subject to jitter from breathing or other motion were removed by evaluating a lag-1 difference over sequential frames and filtering the frames falling above an empirically determined threshold. At this stage an instability resulting from hardware was identified that necessitated the removal of the first 64 frames from each block of 100 from both data-sets. For the awake and anesthetised measurements respectively this left a total number of frames $N_f$ of 8041 and 7679 frames. 

A clutter filter was then used to separate the static or quasi-static signals corresponding to tissue from those originating from blood which was accomplished using a singular value decomposition (SVD)~\cite{demene2015spatiotemporal}. First, both acquisitions were reshaped into a 2D matrix with dimensions $N_{\omega}\times N_t \times N_e, N_f)$ where $N_t$ is the number of transmissions and $N_e$ is the number of elements. Next, the SVD was evaluated over this full set of frames and the data above a manually determined threshold cut-off. For both data-sets this threshold was set at 65$\%$ of the singular values (e.g., for the anesthetised dataset the first 5226 values were discarded). It should be noted that this spatiotemporal filtering is typically applied after image reconstruction, however, we found no significant difference in applying it to our data as a pre-processing step. After this pre-processing our data consists of a 4D tensor $v$ with dimensions $(N_{\omega},N_t,N_e,N_f)$.

\subsection{Acoustic Model}
Conventional ultrasound images are reconstructed geometrically. Simple, non-diffracting, spherical or planar fields are transmitted allowing for images to be reconstructed using only knowledge of the sensor positions and the speed of sound in the propagation medium. The use of a coding-mask and waveguide prohibits this with cUSi as the wave-field evolves in a complex manner with propagation. 

Instead we assume that our data $v$ can be linearly related to a 3D image $u$, via a matrix-vector multiplication 
\begin{equation}
   v = Au.
   \label{yAx}     
\end{equation}
Here we assume the tensors $u$ and $v$ are vectorised and the matrix $A$ contains the pulse-echo impulse response to our transmission scheme for each voxel in the imaging medium. The matrix $A$ has number of rows equal to $N_{\omega}N_tN_e$, and number of columns equal to the number of voxels in the reconstructed image. Reconstructing an image requires precise knowledge of this matrix $A$. Additionally, as the dimensions for $A$ in this work are $8\times10^6\times5\times10^5$ it occupies over 10 Tb in memory so cannot be stored. As such we construct and apply it sequentially.

To construct $A$, first we perform a one-time experimental calibration of the forward field of each element. The probe with the encoding mask and waveguide attached was mounted in a custom-built scanning tank with a three-axis computer-controlled positioning system formed from 3 translation stages X-LSM200B, X-LSM100B, and X-LDA075A (Zaber, Vancouver, Canada) and driven using the ultrasound research system. The impulse response for each element was measured over a 12$\times$12 mm plane parallel to the end of the waveguide with a 40 $\mu m$ spacing using a broadband 0.2 mm needle hydrophone (Precision Acoustics, Dorchester). Signals were sampled using a programmable analog-to-digital converter (M4i.4450-x8, Spectrum, Germany) with 14 bits per sample and a 250-MHz sampling rate. The forward field for each of the 64 elements was measured in a single scan. To improve SNR, signals were averaged 8 times at each position, to reduce the amount of averaging required the probe was driven using the Hadamard encoded synthetic aperture scheme employed for imaging. To measure the probe's full bandwidth it was driven using a 32 Vpp 32 ns impulse, however, to further reduce the required averaging we used temporal Golay codes following Bae et al~\cite{bae2002orthogonal}. After measurement both the temporal and spatial encoding were deconvolved. The calibration took approximately 6 hours and generated a set of 64 spatial temporal measurements $p_{N_e}(x,y,z_s,t)$, where $z_s$ is the plane over which the calibration was performed. These wavefields were then mapped onto the 128 frequencies $N_{\omega}$ that comprised the experimental data. The same calibration was used to reconstruct both data-sets, acquired on separate days, demonstrating that the calibration is robust to repetition. Maximum amplitude projections through each of the elements forward fields are included as Supplementary Movie 3. 

The propagation is assumed to be linear, allowing us to predict the forward field at a location $(x,y,z_s)$ for a frequency $n_{\omega}$ for any arbitrary spatial apodisation vector $H_{N_e}$ and temporal delay vector $T_{N_e}$ applied on transmission as a simple weighted summation  
\begin{equation}
   p(x,y,z_s,n_{\omega}) = \sum_{n_e=1}^{N_e} H_{n_e}p_{n_e}(x,y,z_s,n_{\omega})e^{-i2\pi n_{\omega}T_{n_e}}
   \label{tempconv}     
\end{equation}
In our case we apply no delays to any of the elements and simply use a 64$\times$64 apodisation matrix for our transmissions $H_{N_e}^{N_t}$. So the forward field for any of the transmissions $n_t$ can be calculated simply as 
\begin{equation}
   p_{n_t}(x,y,z_s,n_{\omega}) = \sum_{n_e=1}^{N_e} H^{n_t}_{n_e}p_{n_e}(x,y,z_s,n_{\omega})
   \label{tempconv}     
\end{equation}
Additionally, reciprocity means that we can assume each element in the matrix probe behaves identically on transmit as on receive. Therefore, the pulse-echo impulse response for an element $n_e$ for a transmission $n_t$ for a position in the calibration plane can be calculated via temporal convolution reducing to a multiplication in the frequency domain. 

This allows us to evaluate $u = A^{H}v$ for any voxel in the calibration plane $z_s$ for a frame $n_f$ via

\begin{dmath}
   u(x,y,z_s,n_f) = \sum_{n_{\omega}=1}^{N_{\omega}}\sum_{n_e=1}^{N_e}\sum_{n_t=1}^{N_t} \\ 
   \{p_{n_e}(x,y,z_s,n_{\omega})p_{n_t}(x,y,z_s,n_{\omega})\}^{*}\cdot v(n_{\omega},n_t,n_e,n_f), 
   \label{planarXA}     
\end{dmath}

and, similarly, $v = Au$ can be evaluated as 
\begin{dmath}
   v(n_{\omega},n_t,n_e,n_f) = \int\int \{p_{n_e}(x,y,z_s,n_{\omega})\\
   p_{n_t}(x,y,z_s,n_{\omega})\} u(x,y,z_s,n_f) dxdy. 
   \label{planeYA}     
\end{dmath}

To reconstruct other depths we use the angular spectrum method (ASM) to project the individual forward fields. For an acoustic field over a 2D plane $p(x,y,z_s)$ at a frequency $f$ this calculates the field over a parallel plane at a depth $z_d = z_s + d$ as
\begin{equation}
    p(x,y,z_d) = \mathcal{F}_{k_x k_y}^{-1}\{\mathcal{F}_{xy}\{p(x,y,z_s)\}\cdot H(k_x,k_y,d)\}
    \label{ASM}
\end{equation}
Here $\mathcal{F}_{xy}$ and $\mathcal{F}_{k_x k_y}^{-1}$ denote the 2D discrete Fourier and inverse Fourier transforms respectively and $H(k_x,k_y,d)$ is a propagator function given by 
\begin{equation}
  H(k_x,k_y,d) = 
    \frac{e^{-id\sqrt{k_x^{2}+k_y^{2}-k^{2}}}}{i\sqrt{k_x^{2}+k_y^{2}-k^{2}}}, 
    \label{propagator}
\end{equation}
where $k$ is the wavenumber. Following Zeng and McGough~\cite{Zeng2008} we apply an angular cutoff to the propagator 
\begin{equation}
   k_c = k\sqrt{\frac{D^{2}/2}{D^{2}/2+d^{2}}},
\end{equation}
to eliminate the under-sampled and evanescent spatial frequencies, where $D$ is the lateral dimension of the propagated field. 

Practically for reconstruction it is necessary to discretise our imaging domain. For the images in this work a 40$\times$40$\times$40 $\mu$m spacing was used matching the calibration measurement (to avoid resampling) and corresponding to a spacing of $\frac{\lambda}{2.4}$. The image domain for the images in Fig. \ref{Figure_2}  was 6.8$\times$9.2$\times$8 mm while for the images in Fig.\ \ref{SFig_2} it was 7$\times$6.2$\times$7.2 mm. Finally, calculation of $u = A^Hv$ and $v = Au$ proceeds by first calculating the field for each transmission over the current depth using Eq.\ \ref{tempconv}. Next, $A^Hv$ or $Au$ are evaluated for the current depth using Eq.\ \ref{planarXA} or \ref{planeYA}. Finally, each of the individual elements forward fields are stepped to the next depth using Eq.\ \ref{ASM} and the process is repeated for this depth. 

\subsection{Image Reconstruction.}
Our image reconstruction was formalised in a set of linear equations $v = Au$ allowing a variety of different solvers to be used for reconstruction. We compared 3 different methods. 

The first method was the least squares estimate, which finds an image minimising the square error between the modelled $Au$ and the measured data $v$ as follows:
\begin{equation}
   u' = \mathrm{argmin}_{u'}\lVert v-Au'\rVert^{2}_{2}.
   \label{LSQR}
\end{equation}
This was implemented using the LSMR algorithm~\cite{fong2011lsmr}, similar to the LSQR algorithm~\cite{paige1982lsqr} which is regularly used for large-scale sparse systems, however, safer to use in the case of early termination that was necessary here. Coronal maximum amplitude projections through the complete PDI as well as a fixed 800 $\mu m$ slice of the volume for increasing LSMR iteration are shown in Fig.\ \ref{SFig_5}. The underlying vasculature can be clearly resolved in each image, however, for early iterations there are large variations in the vessel intensity and background caused by spatial variation in the sensitivity of the underlying model~\cite{berthon2018spatiotemporal}. These are compensated for by later iterations and beyond iteration 8 there is minimal change in the reconstructed volume aside from a gradual loss of contrast as the solver fits the noise in the data and errors in the underlying model. This has been reported previously~\cite{Kruizinga2017}, however, occurs more rapidly for the in-vivo data reconstructed here. We attribute this more severe noise amplification to the lower SNR present in ultrafast imaging compared to the sparse, rigid, objects imaged in previous work. 

To try to alleviate the loss of contrast we also tested a sparsity promoting reconstruction method choosing the iterative two-step shrinkage/thresholding algorithm (TwIST~\cite{bioucas2007new}) which minimises the following cost-function 
\begin{equation}
   u' = \mathrm{argmin}_{u'}\lVert v-Au'\rVert^{2}_{2} + \lambda\lVert u'\rVert_1.
   \label{Twist}
\end{equation}
Here $\lambda$ is a dimensionless scalar weighting the regularisation, and $\lVert u'\rVert_1$ is the l1-norm. The regularisation parameter was chosen empirically as $\lambda = 0.2 \mathrm{max}(A^Hv)$. After convergence of the initial algorithm we applied a debiasing step to the non-zero elements of $u$ using a conjugate gradient method following Figueiredo et al~\cite{figueiredo2007gradient}. However, TwIST with debiasing was found to be effectively equivalent with the early iterations of LSMR (Fig.\ \ref{SFig_5}B) which we attribute to the lack of spatial sparsity in the individual frames used to form each PDI.

Both algorithms share two drawbacks. First, the calibration of the model $A$ was performed using a 200 $\mu m$ needle hydrophone due to the poor SNR of smaller sensors. This is larger than the acoustic wavelength of the matrix probe and, as a result, the higher-spatial frequencies of element forward fields are under-estimated~\cite{wear2022spatiotemporal}. This results in a model that is less divergent than the actual experimental field and biases against vessels in the periphery of the volume. Second, given the large-size of the model matrix $A$ iterative approaches that require it to be repeatedly constructed and applied are time-consuming to evaluate. These drawbacks motivated the use of a third approach which we applied for the results presented in both Fig.\ 2 and Fig.\ \ref{SFig_2}. Taking inspiration from observations on one-bit time-reversal where the amplitude information was superfluous~\cite{derode1999ultrasonic} we modified $A$ to contain only the phase-information (i.e., $A_{new} = e^{arg(A)}$ for each voxel and reconstructed images using a matched filter, i.e., $u = A_{new}^Hv$. A comparison of this method with TwIST and LSMR is shown in Fig.\ \ref{SFig_5}B. Taking only the phase of the model re-weights the sensitivity to unity for each voxel. This re-weighting is able to partly compensate for the directivity errors introduced by the measurement method at the expense of diminished contrast. In addition this approach is significantly faster requiring only a single evaluation of $A^Hv$.

Once the full set of frames $N_f$ were reconstructed for both data-sets, we formed the PDI's from a simple summation of the power in each individual voxel as $PDI(x,y,z) = \sum_{n_f=1}^{N_f} |u(x,y,z,n_f)|^2$. The colour Doppler images (CDI) shown in Fig. 2B and S2B were calculated as $arg(\sum_{n_f=1}^{N_f-1}u(x,y,z,n_f)u(x,y,z,n_f+1)^{*})$. However, we found the CDI were noisy when we applied this to volumes reconstructed from separate groups of 64 transmissions. Instead we reconstructed a separate sub-frame for each set of 8 transmissions. A running average was then evaluated over these sub-frames to create a stack of $8N_f$ images each formed from 64 orthogonal transmissions with a lag of 8 transmissions between images. The CDI processing was then applied to this new, larger, image stack. 

Here we have shown that with a simple, quick, reconstruction method we are still able to form high-resolution 3D images of the vasculature. In the future there is significant scope to apply more advanced reconstruction methods that exploit statistical independence of the signals in neighbouring voxels~\cite{dogan2021multiple,bar2018sushi}. Additionally, the directivity and other errors associated with the experimental calibration of $A$ used here could be eliminated with the use of blind calibration methods~\cite{van2018calibration} or by deconvolution of the directivity response~\cite{wear2022spatiotemporal}. 

\subsection{Image Resolution.}
We approximated our image resolution with two distinct approaches. First, by calculating the decay rate of the lateral and axial correlations of our system matrix $A$ at three different depths. Second, by approximating the length scale on which we can distinguish separate cortical vessels over a fixed depth. These are illustrated in Fig. \ref{SFig_6}. 

The correlations were evaluated at depths corresponding to approximately 1.5 mm, 4.7 mm and 7.9 mm inside the mouse brain in the experimental data. In each case a single voxel was selected and the magnitude of the correlation coefficient of this voxel with each voxel of the system matrix $A$ over a 2$\times$2$\times$1 mm region with a 40 $\mu m$ spacing was calculated. The results demonstrate that both the lateral and axial resolution decrease with imaging depth, however, this occurs more slowly for the axial resolution which is additionally higher than the lateral resolution. This higher axial resolution is commonly the case for ultrasound imaging. Higher side-lobes are seen in the axial direction, however, these are largely confined to the lateral position of the voxel-of-interest (i.e., the side-lobes are not spherically symmetric). By zooming in on a fixed axial slice of the reconstructed PDI we can see that this correlation-based approximation of the lateral resolution corresponds well with the scale of structures reconstructed from the experimental data. With vessels in the cortex being resolvable, as defined by the Rayleigh criterion~\cite{Antipa2018}, with a spacing of less than 200 $\mu m$. 
\bibliography{Manuscript.bib}

\begin{thebibliography}{64}%
\makeatletter
\providecommand \@ifxundefined [1]{%
 \@ifx{#1\undefined}
}%
\providecommand \@ifnum [1]{%
 \ifnum #1\expandafter \@firstoftwo
 \else \expandafter \@secondoftwo
 \fi
}%
\providecommand \@ifx [1]{%
 \ifx #1\expandafter \@firstoftwo
 \else \expandafter \@secondoftwo
 \fi
}%
\providecommand \natexlab [1]{#1}%
\providecommand \enquote  [1]{``#1''}%
\providecommand \bibnamefont  [1]{#1}%
\providecommand \bibfnamefont [1]{#1}%
\providecommand \citenamefont [1]{#1}%
\providecommand \href@noop [0]{\@secondoftwo}%
\providecommand \href [0]{\begingroup \@sanitize@url \@href}%
\providecommand \@href[1]{\@@startlink{#1}\@@href}%
\providecommand \@@href[1]{\endgroup#1\@@endlink}%
\providecommand \@sanitize@url [0]{\catcode `\\12\catcode `\$12\catcode
  `\&12\catcode `\#12\catcode `\^12\catcode `\_12\catcode `\%12\relax}%
\providecommand \@@startlink[1]{}%
\providecommand \@@endlink[0]{}%
\providecommand \url  [0]{\begingroup\@sanitize@url \@url }%
\providecommand \@url [1]{\endgroup\@href {#1}{\urlprefix }}%
\providecommand \urlprefix  [0]{URL }%
\providecommand \Eprint [0]{\href }%
\providecommand \doibase [0]{https://doi.org/}%
\providecommand \selectlanguage [0]{\@gobble}%
\providecommand \bibinfo  [0]{\@secondoftwo}%
\providecommand \bibfield  [0]{\@secondoftwo}%
\providecommand \translation [1]{[#1]}%
\providecommand \BibitemOpen [0]{}%
\providecommand \bibitemStop [0]{}%
\providecommand \bibitemNoStop [0]{.\EOS\space}%
\providecommand \EOS [0]{\spacefactor3000\relax}%
\providecommand \BibitemShut  [1]{\csname bibitem#1\endcsname}%
\let\auto@bib@innerbib\@empty
\bibitem [{\citenamefont {Tanter}\ and\ \citenamefont
  {Fink}(2014)}]{Tanter2014}%
  \BibitemOpen
  \bibfield  {author} {\bibinfo {author} {\bibfnamefont {M.}~\bibnamefont
  {Tanter}}\ and\ \bibinfo {author} {\bibfnamefont {M.}~\bibnamefont {Fink}},\
  }\bibfield  {title} {\bibinfo {title} {Ultrafast imaging in biomedical
  ultrasound},\ }\href {https://doi.org/10.1109/TUFFC.2014.2882} {\bibfield
  {journal} {\bibinfo  {journal} {IEEE Transactions on Ultrasonics,
  Ferroelectrics, and Frequency Control}\ }\textbf {\bibinfo {volume} {61}},\
  \bibinfo {pages} {102} (\bibinfo {year} {2014})}\BibitemShut {NoStop}%
\bibitem [{\citenamefont {Couade}(2016)}]{Couade2016}%
  \BibitemOpen
  \bibfield  {author} {\bibinfo {author} {\bibfnamefont {M.}~\bibnamefont
  {Couade}},\ }\bibfield  {title} {\bibinfo {title} {The advent of ultrafast
  ultrasound in vascular imaging: a review},\ }\href
  {https://doi.org/10.2147/jvd.s68045} {\bibfield  {journal} {\bibinfo
  {journal} {Journal of Vascular Diagnostics and Interventions}\ ,\ \bibinfo
  {pages} {9}} (\bibinfo {year} {2016})}\BibitemShut {NoStop}%
\bibitem [{\citenamefont {Sebag}\ \emph {et~al.}(2010)\citenamefont {Sebag},
  \citenamefont {Vaillant-Lombard}, \citenamefont {Berbis}, \citenamefont
  {Griset}, \citenamefont {Henry}, \citenamefont {Petit},\ and\ \citenamefont
  {Oliver}}]{Sebag2010}%
  \BibitemOpen
  \bibfield  {author} {\bibinfo {author} {\bibfnamefont {F.}~\bibnamefont
  {Sebag}}, \bibinfo {author} {\bibfnamefont {J.}~\bibnamefont
  {Vaillant-Lombard}}, \bibinfo {author} {\bibfnamefont {J.}~\bibnamefont
  {Berbis}}, \bibinfo {author} {\bibfnamefont {V.}~\bibnamefont {Griset}},
  \bibinfo {author} {\bibfnamefont {J.~F.}\ \bibnamefont {Henry}}, \bibinfo
  {author} {\bibfnamefont {P.}~\bibnamefont {Petit}},\ and\ \bibinfo {author}
  {\bibfnamefont {C.}~\bibnamefont {Oliver}},\ }\bibfield  {title} {\bibinfo
  {title} {Shear wave elastography: A new ultrasound imaging mode for the
  differential diagnosis of benign and malignant thyroid nodules},\ }\href
  {https://doi.org/10.1210/jc.2010-0766} {\bibfield  {journal} {\bibinfo
  {journal} {Journal of Clinical Endocrinology and Metabolism}\ }\textbf
  {\bibinfo {volume} {95}},\ \bibinfo {pages} {5281} (\bibinfo {year}
  {2010})}\BibitemShut {NoStop}%
\bibitem [{\citenamefont {Vappou}\ \emph {et~al.}(2010)\citenamefont {Vappou},
  \citenamefont {Luo},\ and\ \citenamefont {Konofagou}}]{Vappou2010}%
  \BibitemOpen
  \bibfield  {author} {\bibinfo {author} {\bibfnamefont {J.}~\bibnamefont
  {Vappou}}, \bibinfo {author} {\bibfnamefont {J.}~\bibnamefont {Luo}},\ and\
  \bibinfo {author} {\bibfnamefont {E.~E.}\ \bibnamefont {Konofagou}},\
  }\bibfield  {title} {\bibinfo {title} {Pulse wave imaging for noninvasive and
  quantitative measurement of arterial stiffness in vivo},\ }\href
  {https://doi.org/10.1038/ajh.2009.272} {\bibfield  {journal} {\bibinfo
  {journal} {American Journal of Hypertension}\ }\textbf {\bibinfo {volume}
  {23}},\ \bibinfo {pages} {393} (\bibinfo {year} {2010})}\BibitemShut
  {NoStop}%
\bibitem [{\citenamefont {Errico}\ \emph {et~al.}(2015)\citenamefont {Errico},
  \citenamefont {Pierre}, \citenamefont {Pezet}, \citenamefont {Desailly},
  \citenamefont {Lenkei}, \citenamefont {Couture},\ and\ \citenamefont
  {Tanter}}]{Errico2015}%
  \BibitemOpen
  \bibfield  {author} {\bibinfo {author} {\bibfnamefont {C.}~\bibnamefont
  {Errico}}, \bibinfo {author} {\bibfnamefont {J.}~\bibnamefont {Pierre}},
  \bibinfo {author} {\bibfnamefont {S.}~\bibnamefont {Pezet}}, \bibinfo
  {author} {\bibfnamefont {Y.}~\bibnamefont {Desailly}}, \bibinfo {author}
  {\bibfnamefont {Z.}~\bibnamefont {Lenkei}}, \bibinfo {author} {\bibfnamefont
  {O.}~\bibnamefont {Couture}},\ and\ \bibinfo {author} {\bibfnamefont
  {M.}~\bibnamefont {Tanter}},\ }\bibfield  {title} {\bibinfo {title}
  {Ultrafast ultrasound localization microscopy for deep super-resolution
  vascular imaging},\ }\href {https://doi.org/10.1038/nature16066} {\bibfield
  {journal} {\bibinfo  {journal} {Nature}\ }\textbf {\bibinfo {volume} {527}},\
  \bibinfo {pages} {499} (\bibinfo {year} {2015})}\BibitemShut {NoStop}%
\bibitem [{\citenamefont {Christensen-Jeffries}\ \emph
  {et~al.}(2020)\citenamefont {Christensen-Jeffries}, \citenamefont {Couture},
  \citenamefont {Dayton}, \citenamefont {Eldar}, \citenamefont {Hynynen},
  \citenamefont {Kiessling}, \citenamefont {O'Reilly}, \citenamefont {Pinton},
  \citenamefont {Schmitz}, \citenamefont {Tang}, \citenamefont {Tanter},\ and\
  \citenamefont {van Sloun}}]{Christensen2020}%
  \BibitemOpen
  \bibfield  {author} {\bibinfo {author} {\bibfnamefont {K.}~\bibnamefont
  {Christensen-Jeffries}}, \bibinfo {author} {\bibfnamefont {O.}~\bibnamefont
  {Couture}}, \bibinfo {author} {\bibfnamefont {P.~A.}\ \bibnamefont {Dayton}},
  \bibinfo {author} {\bibfnamefont {Y.~C.}\ \bibnamefont {Eldar}}, \bibinfo
  {author} {\bibfnamefont {K.}~\bibnamefont {Hynynen}}, \bibinfo {author}
  {\bibfnamefont {F.}~\bibnamefont {Kiessling}}, \bibinfo {author}
  {\bibfnamefont {M.}~\bibnamefont {O'Reilly}}, \bibinfo {author}
  {\bibfnamefont {G.~F.}\ \bibnamefont {Pinton}}, \bibinfo {author}
  {\bibfnamefont {G.}~\bibnamefont {Schmitz}}, \bibinfo {author} {\bibfnamefont
  {M.~X.}\ \bibnamefont {Tang}}, \bibinfo {author} {\bibfnamefont
  {M.}~\bibnamefont {Tanter}},\ and\ \bibinfo {author} {\bibfnamefont {R.~J.}\
  \bibnamefont {van Sloun}},\ }\bibfield  {title} {\bibinfo {title}
  {Super-resolution ultrasound imaging},\ }\href
  {https://doi.org/10.1016/j.ultrasmedbio.2019.11.013} {\bibfield  {journal}
  {\bibinfo  {journal} {Ultrasound in Medicine and Biology}\ }\textbf {\bibinfo
  {volume} {46}},\ \bibinfo {pages} {865} (\bibinfo {year} {2020})}\BibitemShut
  {NoStop}%
\bibitem [{\citenamefont {Bercoff}\ \emph {et~al.}(2011)\citenamefont
  {Bercoff}, \citenamefont {Montaldo}, \citenamefont {Loupas}, \citenamefont
  {Savery}, \citenamefont {Mézière}, \citenamefont {Fink},\ and\
  \citenamefont {Tanter}}]{Bercoff2011}%
  \BibitemOpen
  \bibfield  {author} {\bibinfo {author} {\bibfnamefont {J.}~\bibnamefont
  {Bercoff}}, \bibinfo {author} {\bibfnamefont {G.}~\bibnamefont {Montaldo}},
  \bibinfo {author} {\bibfnamefont {T.}~\bibnamefont {Loupas}}, \bibinfo
  {author} {\bibfnamefont {D.}~\bibnamefont {Savery}}, \bibinfo {author}
  {\bibfnamefont {F.}~\bibnamefont {Mézière}}, \bibinfo {author}
  {\bibfnamefont {M.}~\bibnamefont {Fink}},\ and\ \bibinfo {author}
  {\bibfnamefont {M.}~\bibnamefont {Tanter}},\ }\bibfield  {title} {\bibinfo
  {title} {Ultrafast compound doppler imaging: Providing full blood flow
  characterization},\ }\href {https://doi.org/10.1109/TUFFC.2011.1780}
  {\bibfield  {journal} {\bibinfo  {journal} {IEEE Transactions on Ultrasonics,
  Ferroelectrics, and Frequency Control}\ }\textbf {\bibinfo {volume} {58}},\
  \bibinfo {pages} {134} (\bibinfo {year} {2011})}\BibitemShut {NoStop}%
\bibitem [{\citenamefont {Mace}\ \emph {et~al.}(2011)\citenamefont {Mace},
  \citenamefont {Montaldo}, \citenamefont {Cohen}, \citenamefont {Baulac},
  \citenamefont {Fink},\ and\ \citenamefont {Tanter}}]{Mace2011}%
  \BibitemOpen
  \bibfield  {author} {\bibinfo {author} {\bibfnamefont {E.}~\bibnamefont
  {Mace}}, \bibinfo {author} {\bibfnamefont {G.}~\bibnamefont {Montaldo}},
  \bibinfo {author} {\bibfnamefont {I.}~\bibnamefont {Cohen}}, \bibinfo
  {author} {\bibfnamefont {M.}~\bibnamefont {Baulac}}, \bibinfo {author}
  {\bibfnamefont {M.}~\bibnamefont {Fink}},\ and\ \bibinfo {author}
  {\bibfnamefont {M.}~\bibnamefont {Tanter}},\ }\bibfield  {title} {\bibinfo
  {title} {Functional ultrasound imaging of the brain},\ }\href
  {https://doi.org/10.1038/nmeth.1641} {\bibfield  {journal} {\bibinfo
  {journal} {Nature Methods}\ }\textbf {\bibinfo {volume} {8}},\ \bibinfo
  {pages} {662} (\bibinfo {year} {2011})}\BibitemShut {NoStop}%
\bibitem [{\citenamefont {Deffieux}\ \emph {et~al.}(2021)\citenamefont
  {Deffieux}, \citenamefont {Demené},\ and\ \citenamefont
  {Tanter}}]{Deffieux2021}%
  \BibitemOpen
  \bibfield  {author} {\bibinfo {author} {\bibfnamefont {T.}~\bibnamefont
  {Deffieux}}, \bibinfo {author} {\bibfnamefont {C.}~\bibnamefont {Demené}},\
  and\ \bibinfo {author} {\bibfnamefont {M.}~\bibnamefont {Tanter}},\
  }\bibfield  {title} {\bibinfo {title} {Functional ultrasound imaging: A new
  imaging modality for neuroscience},\ }\href
  {https://doi.org/10.1016/j.neuroscience.2021.03.005} {\bibfield  {journal}
  {\bibinfo  {journal} {Neuroscience}\ }\textbf {\bibinfo {volume} {474}},\
  \bibinfo {pages} {110} (\bibinfo {year} {2021})}\BibitemShut {NoStop}%
\bibitem [{\citenamefont {Renaudin}\ \emph {et~al.}(2022)\citenamefont
  {Renaudin}, \citenamefont {Demené}, \citenamefont {Dizeux}, \citenamefont
  {Ialy-Radio}, \citenamefont {Pezet},\ and\ \citenamefont
  {Tanter}}]{Renaudin2022}%
  \BibitemOpen
  \bibfield  {author} {\bibinfo {author} {\bibfnamefont {N.}~\bibnamefont
  {Renaudin}}, \bibinfo {author} {\bibfnamefont {C.}~\bibnamefont {Demené}},
  \bibinfo {author} {\bibfnamefont {A.}~\bibnamefont {Dizeux}}, \bibinfo
  {author} {\bibfnamefont {N.}~\bibnamefont {Ialy-Radio}}, \bibinfo {author}
  {\bibfnamefont {S.}~\bibnamefont {Pezet}},\ and\ \bibinfo {author}
  {\bibfnamefont {M.}~\bibnamefont {Tanter}},\ }\bibfield  {title} {\bibinfo
  {title} {Functional ultrasound localization microscopy reveals brain-wide
  neurovascular activity on a microscopic scale},\ }\href
  {https://doi.org/10.1038/s41592-022-01549-5} {\bibfield  {journal} {\bibinfo
  {journal} {Nature Methods}\ }\textbf {\bibinfo {volume} {19}},\ \bibinfo
  {pages} {1004} (\bibinfo {year} {2022})}\BibitemShut {NoStop}%
\bibitem [{\citenamefont {Provost}\ \emph {et~al.}(2014)\citenamefont
  {Provost}, \citenamefont {Papadacci}, \citenamefont {Arango}, \citenamefont
  {Imbault}, \citenamefont {Fink}, \citenamefont {Gennisson}, \citenamefont
  {Tanter},\ and\ \citenamefont {Pernot}}]{provost20143d}%
  \BibitemOpen
  \bibfield  {author} {\bibinfo {author} {\bibfnamefont {J.}~\bibnamefont
  {Provost}}, \bibinfo {author} {\bibfnamefont {C.}~\bibnamefont {Papadacci}},
  \bibinfo {author} {\bibfnamefont {J.~E.}\ \bibnamefont {Arango}}, \bibinfo
  {author} {\bibfnamefont {M.}~\bibnamefont {Imbault}}, \bibinfo {author}
  {\bibfnamefont {M.}~\bibnamefont {Fink}}, \bibinfo {author} {\bibfnamefont
  {J.-L.}\ \bibnamefont {Gennisson}}, \bibinfo {author} {\bibfnamefont
  {M.}~\bibnamefont {Tanter}},\ and\ \bibinfo {author} {\bibfnamefont
  {M.}~\bibnamefont {Pernot}},\ }\bibfield  {title} {\bibinfo {title} {3d
  ultrafast ultrasound imaging in vivo},\ }\href@noop {} {\bibfield  {journal}
  {\bibinfo  {journal} {Physics in Medicine \& Biology}\ }\textbf {\bibinfo
  {volume} {59}},\ \bibinfo {pages} {L1} (\bibinfo {year} {2014})}\BibitemShut
  {NoStop}%
\bibitem [{\citenamefont {Petrusca}\ \emph {et~al.}(2017)\citenamefont
  {Petrusca}, \citenamefont {Varray}, \citenamefont {Souchon}, \citenamefont
  {Bernard}, \citenamefont {Chapelon}, \citenamefont {Liebgott}, \citenamefont
  {N'Djin},\ and\ \citenamefont {Viallon}}]{petrusca2017new}%
  \BibitemOpen
  \bibfield  {author} {\bibinfo {author} {\bibfnamefont {L.}~\bibnamefont
  {Petrusca}}, \bibinfo {author} {\bibfnamefont {F.}~\bibnamefont {Varray}},
  \bibinfo {author} {\bibfnamefont {R.}~\bibnamefont {Souchon}}, \bibinfo
  {author} {\bibfnamefont {A.}~\bibnamefont {Bernard}}, \bibinfo {author}
  {\bibfnamefont {J.-Y.}\ \bibnamefont {Chapelon}}, \bibinfo {author}
  {\bibfnamefont {H.}~\bibnamefont {Liebgott}}, \bibinfo {author}
  {\bibfnamefont {W.~A.}\ \bibnamefont {N'Djin}},\ and\ \bibinfo {author}
  {\bibfnamefont {M.}~\bibnamefont {Viallon}},\ }\bibfield  {title} {\bibinfo
  {title} {A new high channels density ultrasound platform for advanced 4d
  cardiac imaging},\ }in\ \href@noop {} {\emph {\bibinfo {booktitle} {2017 IEEE
  International Ultrasonics Symposium (IUS)}}}\ (\bibinfo {organization}
  {IEEE},\ \bibinfo {year} {2017})\ pp.\ \bibinfo {pages} {1--4}\BibitemShut
  {NoStop}%
\bibitem [{\citenamefont {Chavignon}\ \emph {et~al.}(2021)\citenamefont
  {Chavignon}, \citenamefont {Heiles}, \citenamefont {Hingot}, \citenamefont
  {Orset}, \citenamefont {Vivien},\ and\ \citenamefont
  {Couture}}]{chavignon20213d}%
  \BibitemOpen
  \bibfield  {author} {\bibinfo {author} {\bibfnamefont {A.}~\bibnamefont
  {Chavignon}}, \bibinfo {author} {\bibfnamefont {B.}~\bibnamefont {Heiles}},
  \bibinfo {author} {\bibfnamefont {V.}~\bibnamefont {Hingot}}, \bibinfo
  {author} {\bibfnamefont {C.}~\bibnamefont {Orset}}, \bibinfo {author}
  {\bibfnamefont {D.}~\bibnamefont {Vivien}},\ and\ \bibinfo {author}
  {\bibfnamefont {O.}~\bibnamefont {Couture}},\ }\bibfield  {title} {\bibinfo
  {title} {3d transcranial ultrasound localization microscopy in the rat brain
  with a multiplexed matrix probe},\ }\href@noop {} {\bibfield  {journal}
  {\bibinfo  {journal} {IEEE Transactions on Biomedical Engineering}\ }\textbf
  {\bibinfo {volume} {69}},\ \bibinfo {pages} {2132} (\bibinfo {year}
  {2021})}\BibitemShut {NoStop}%
\bibitem [{\citenamefont {Heiles}\ \emph {et~al.}(2019)\citenamefont {Heiles},
  \citenamefont {Correia}, \citenamefont {Hingot}, \citenamefont {Pernot},
  \citenamefont {Provost}, \citenamefont {Tanter},\ and\ \citenamefont
  {Couture}}]{heiles2019ultrafast}%
  \BibitemOpen
  \bibfield  {author} {\bibinfo {author} {\bibfnamefont {B.}~\bibnamefont
  {Heiles}}, \bibinfo {author} {\bibfnamefont {M.}~\bibnamefont {Correia}},
  \bibinfo {author} {\bibfnamefont {V.}~\bibnamefont {Hingot}}, \bibinfo
  {author} {\bibfnamefont {M.}~\bibnamefont {Pernot}}, \bibinfo {author}
  {\bibfnamefont {J.}~\bibnamefont {Provost}}, \bibinfo {author} {\bibfnamefont
  {M.}~\bibnamefont {Tanter}},\ and\ \bibinfo {author} {\bibfnamefont
  {O.}~\bibnamefont {Couture}},\ }\bibfield  {title} {\bibinfo {title}
  {Ultrafast 3d ultrasound localization microscopy using a 32$\times$32 matrix
  array},\ }\href@noop {} {\bibfield  {journal} {\bibinfo  {journal} {IEEE
  Transactions on Medical Imaging}\ }\textbf {\bibinfo {volume} {38}},\
  \bibinfo {pages} {2005} (\bibinfo {year} {2019})}\BibitemShut {NoStop}%
\bibitem [{\citenamefont {Rabut}\ \emph {et~al.}(2019)\citenamefont {Rabut},
  \citenamefont {Correia}, \citenamefont {Finel}, \citenamefont {Pezet},
  \citenamefont {Pernot}, \citenamefont {Deffieux},\ and\ \citenamefont
  {Tanter}}]{Rabut2019}%
  \BibitemOpen
  \bibfield  {author} {\bibinfo {author} {\bibfnamefont {C.}~\bibnamefont
  {Rabut}}, \bibinfo {author} {\bibfnamefont {M.}~\bibnamefont {Correia}},
  \bibinfo {author} {\bibfnamefont {V.}~\bibnamefont {Finel}}, \bibinfo
  {author} {\bibfnamefont {S.}~\bibnamefont {Pezet}}, \bibinfo {author}
  {\bibfnamefont {M.}~\bibnamefont {Pernot}}, \bibinfo {author} {\bibfnamefont
  {T.}~\bibnamefont {Deffieux}},\ and\ \bibinfo {author} {\bibfnamefont
  {M.}~\bibnamefont {Tanter}},\ }\bibfield  {title} {\bibinfo {title} {4d
  functional ultrasound imaging of whole-brain activity in rodents},\ }\href
  {https://doi.org/10.1038/s41592-019-0572-y} {\bibfield  {journal} {\bibinfo
  {journal} {Nature Methods}\ }\textbf {\bibinfo {volume} {16}},\ \bibinfo
  {pages} {994} (\bibinfo {year} {2019})}\BibitemShut {NoStop}%
\bibitem [{\citenamefont {Chen}\ \emph {et~al.}(2017)\citenamefont {Chen},
  \citenamefont {Chen}, \citenamefont {Bera}, \citenamefont {Raghunathan},
  \citenamefont {Shabanimotlagh}, \citenamefont {Noothout}, \citenamefont
  {Chang}, \citenamefont {Ponte}, \citenamefont {Prins}, \citenamefont {Vos}
  \emph {et~al.}}]{Chen2017}%
  \BibitemOpen
  \bibfield  {author} {\bibinfo {author} {\bibfnamefont {C.}~\bibnamefont
  {Chen}}, \bibinfo {author} {\bibfnamefont {Z.}~\bibnamefont {Chen}}, \bibinfo
  {author} {\bibfnamefont {D.}~\bibnamefont {Bera}}, \bibinfo {author}
  {\bibfnamefont {S.~B.}\ \bibnamefont {Raghunathan}}, \bibinfo {author}
  {\bibfnamefont {M.}~\bibnamefont {Shabanimotlagh}}, \bibinfo {author}
  {\bibfnamefont {E.}~\bibnamefont {Noothout}}, \bibinfo {author}
  {\bibfnamefont {Z.-Y.}\ \bibnamefont {Chang}}, \bibinfo {author}
  {\bibfnamefont {J.}~\bibnamefont {Ponte}}, \bibinfo {author} {\bibfnamefont
  {C.}~\bibnamefont {Prins}}, \bibinfo {author} {\bibfnamefont {H.~J.}\
  \bibnamefont {Vos}}, \emph {et~al.},\ }\bibfield  {title} {\bibinfo {title}
  {A front-end asic with receive sub-array beamforming integrated with a
  32$\times$32 pzt matrix transducer for 3-d transesophageal
  echocardiography},\ }\href@noop {} {\bibfield  {journal} {\bibinfo  {journal}
  {IEEE Journal of Solid-State Circuits}\ }\textbf {\bibinfo {volume} {52}},\
  \bibinfo {pages} {994} (\bibinfo {year} {2017})}\BibitemShut {NoStop}%
\bibitem [{\citenamefont {Janjic}\ \emph
  {et~al.}(2018{\natexlab{a}})\citenamefont {Janjic}, \citenamefont {Tan},
  \citenamefont {Daeichin}, \citenamefont {Noothout}, \citenamefont {Chen},
  \citenamefont {Chen}, \citenamefont {Chang}, \citenamefont {Beurskens},
  \citenamefont {Van~Soest}, \citenamefont {Van Der~Steen} \emph
  {et~al.}}]{janjic20182}%
  \BibitemOpen
  \bibfield  {author} {\bibinfo {author} {\bibfnamefont {J.}~\bibnamefont
  {Janjic}}, \bibinfo {author} {\bibfnamefont {M.}~\bibnamefont {Tan}},
  \bibinfo {author} {\bibfnamefont {V.}~\bibnamefont {Daeichin}}, \bibinfo
  {author} {\bibfnamefont {E.}~\bibnamefont {Noothout}}, \bibinfo {author}
  {\bibfnamefont {C.}~\bibnamefont {Chen}}, \bibinfo {author} {\bibfnamefont
  {Z.}~\bibnamefont {Chen}}, \bibinfo {author} {\bibfnamefont {Z.-Y.}\
  \bibnamefont {Chang}}, \bibinfo {author} {\bibfnamefont {R.~H.}\ \bibnamefont
  {Beurskens}}, \bibinfo {author} {\bibfnamefont {G.}~\bibnamefont
  {Van~Soest}}, \bibinfo {author} {\bibfnamefont {A.~F.}\ \bibnamefont {Van
  Der~Steen}}, \emph {et~al.},\ }\bibfield  {title} {\bibinfo {title} {A 2-d
  ultrasound transducer with front-end asic and low cable count for 3-d
  forward-looking intravascular imaging: Performance and characterization},\
  }\href@noop {} {\bibfield  {journal} {\bibinfo  {journal} {IEEE transactions
  on ultrasonics, ferroelectrics, and frequency control}\ }\textbf {\bibinfo
  {volume} {65}},\ \bibinfo {pages} {1832} (\bibinfo {year}
  {2018}{\natexlab{a}})}\BibitemShut {NoStop}%
\bibitem [{\citenamefont {Sauvage}\ \emph {et~al.}(2019)\citenamefont
  {Sauvage}, \citenamefont {Por{\'e}e}, \citenamefont {Rabut}, \citenamefont
  {F{\'e}rin}, \citenamefont {Flesch}, \citenamefont {Rosinski}, \citenamefont
  {Nguyen-Dinh}, \citenamefont {Tanter}, \citenamefont {Pernot},\ and\
  \citenamefont {Deffieux}}]{sauvage20194d}%
  \BibitemOpen
  \bibfield  {author} {\bibinfo {author} {\bibfnamefont {J.}~\bibnamefont
  {Sauvage}}, \bibinfo {author} {\bibfnamefont {J.}~\bibnamefont {Por{\'e}e}},
  \bibinfo {author} {\bibfnamefont {C.}~\bibnamefont {Rabut}}, \bibinfo
  {author} {\bibfnamefont {G.}~\bibnamefont {F{\'e}rin}}, \bibinfo {author}
  {\bibfnamefont {M.}~\bibnamefont {Flesch}}, \bibinfo {author} {\bibfnamefont
  {B.}~\bibnamefont {Rosinski}}, \bibinfo {author} {\bibfnamefont
  {A.}~\bibnamefont {Nguyen-Dinh}}, \bibinfo {author} {\bibfnamefont
  {M.}~\bibnamefont {Tanter}}, \bibinfo {author} {\bibfnamefont
  {M.}~\bibnamefont {Pernot}},\ and\ \bibinfo {author} {\bibfnamefont
  {T.}~\bibnamefont {Deffieux}},\ }\bibfield  {title} {\bibinfo {title} {4d
  functional imaging of the rat brain using a large aperture row-column
  array},\ }\href@noop {} {\bibfield  {journal} {\bibinfo  {journal} {IEEE
  transactions on medical imaging}\ }\textbf {\bibinfo {volume} {39}},\
  \bibinfo {pages} {1884} (\bibinfo {year} {2019})}\BibitemShut {NoStop}%
\bibitem [{\citenamefont {Wei}\ \emph {et~al.}(2021)\citenamefont {Wei},
  \citenamefont {Wahyulaksana}, \citenamefont {Meijlink}, \citenamefont
  {Ramalli}, \citenamefont {Noothout}, \citenamefont {Verweij}, \citenamefont
  {Boni}, \citenamefont {Kooiman}, \citenamefont {Van Der~Steen}, \citenamefont
  {Tortoli} \emph {et~al.}}]{wei2021high}%
  \BibitemOpen
  \bibfield  {author} {\bibinfo {author} {\bibfnamefont {L.}~\bibnamefont
  {Wei}}, \bibinfo {author} {\bibfnamefont {G.}~\bibnamefont {Wahyulaksana}},
  \bibinfo {author} {\bibfnamefont {B.}~\bibnamefont {Meijlink}}, \bibinfo
  {author} {\bibfnamefont {A.}~\bibnamefont {Ramalli}}, \bibinfo {author}
  {\bibfnamefont {E.}~\bibnamefont {Noothout}}, \bibinfo {author}
  {\bibfnamefont {M.~D.}\ \bibnamefont {Verweij}}, \bibinfo {author}
  {\bibfnamefont {E.}~\bibnamefont {Boni}}, \bibinfo {author} {\bibfnamefont
  {K.}~\bibnamefont {Kooiman}}, \bibinfo {author} {\bibfnamefont {A.~F.}\
  \bibnamefont {Van Der~Steen}}, \bibinfo {author} {\bibfnamefont
  {P.}~\bibnamefont {Tortoli}}, \emph {et~al.},\ }\bibfield  {title} {\bibinfo
  {title} {High frame rate volumetric imaging of microbubbles using a sparse
  array and spatial coherence beamforming},\ }\href@noop {} {\bibfield
  {journal} {\bibinfo  {journal} {IEEE Transactions on Ultrasonics,
  Ferroelectrics, and Frequency Control}\ }\textbf {\bibinfo {volume} {68}},\
  \bibinfo {pages} {3069} (\bibinfo {year} {2021})}\BibitemShut {NoStop}%
\bibitem [{\citenamefont {Ramalli}\ \emph {et~al.}(2022)\citenamefont
  {Ramalli}, \citenamefont {Boni}, \citenamefont {Roux}, \citenamefont
  {Liebgott},\ and\ \citenamefont {Tortoli}}]{Ramalli2022}%
  \BibitemOpen
  \bibfield  {author} {\bibinfo {author} {\bibfnamefont {A.}~\bibnamefont
  {Ramalli}}, \bibinfo {author} {\bibfnamefont {E.}~\bibnamefont {Boni}},
  \bibinfo {author} {\bibfnamefont {E.}~\bibnamefont {Roux}}, \bibinfo {author}
  {\bibfnamefont {H.}~\bibnamefont {Liebgott}},\ and\ \bibinfo {author}
  {\bibfnamefont {P.}~\bibnamefont {Tortoli}},\ }\bibfield  {title} {\bibinfo
  {title} {Design, implementation, and medical applications of 2-d ultrasound
  sparse arrays},\ }\href@noop {} {\bibfield  {journal} {\bibinfo  {journal}
  {IEEE Transactions on Ultrasonics, Ferroelectrics, and Frequency Control}\ }
  (\bibinfo {year} {2022})}\BibitemShut {NoStop}%
\bibitem [{\citenamefont {Favre}\ \emph {et~al.}(2023)\citenamefont {Favre},
  \citenamefont {Pernot}, \citenamefont {Tanter},\ and\ \citenamefont
  {Papadacci}}]{favre2023transcranial}%
  \BibitemOpen
  \bibfield  {author} {\bibinfo {author} {\bibfnamefont {H.}~\bibnamefont
  {Favre}}, \bibinfo {author} {\bibfnamefont {M.}~\bibnamefont {Pernot}},
  \bibinfo {author} {\bibfnamefont {M.}~\bibnamefont {Tanter}},\ and\ \bibinfo
  {author} {\bibfnamefont {C.}~\bibnamefont {Papadacci}},\ }\bibfield  {title}
  {\bibinfo {title} {Transcranial 3d ultrasound localization microscopy using a
  large element matrix array with a multi-lens diffracting layer: an in vitro
  study},\ }\href@noop {} {\bibfield  {journal} {\bibinfo  {journal} {Physics
  in Medicine \& Biology}\ }\textbf {\bibinfo {volume} {68}},\ \bibinfo {pages}
  {075003} (\bibinfo {year} {2023})}\BibitemShut {NoStop}%
\bibitem [{\citenamefont {Szabo}(2004)}]{Szabo2004}%
  \BibitemOpen
  \bibfield  {author} {\bibinfo {author} {\bibfnamefont {T.~L.}\ \bibnamefont
  {Szabo}},\ }\href@noop {} {\emph {\bibinfo {title} {Diagnostic ultrasound
  imaging: inside out}}}\ (\bibinfo  {publisher} {Academic press},\ \bibinfo
  {year} {2004})\BibitemShut {NoStop}%
\bibitem [{\citenamefont {Kruizinga}\ \emph {et~al.}(2017)\citenamefont
  {Kruizinga}, \citenamefont {van~der Meulen}, \citenamefont {Fedjajevs},
  \citenamefont {Mastik}, \citenamefont {Springeling}, \citenamefont {de~Jong},
  \citenamefont {Bosch},\ and\ \citenamefont {Leus}}]{Kruizinga2017}%
  \BibitemOpen
  \bibfield  {author} {\bibinfo {author} {\bibfnamefont {P.}~\bibnamefont
  {Kruizinga}}, \bibinfo {author} {\bibfnamefont {P.}~\bibnamefont {van~der
  Meulen}}, \bibinfo {author} {\bibfnamefont {A.}~\bibnamefont {Fedjajevs}},
  \bibinfo {author} {\bibfnamefont {F.}~\bibnamefont {Mastik}}, \bibinfo
  {author} {\bibfnamefont {G.}~\bibnamefont {Springeling}}, \bibinfo {author}
  {\bibfnamefont {N.}~\bibnamefont {de~Jong}}, \bibinfo {author} {\bibfnamefont
  {J.~G.}\ \bibnamefont {Bosch}},\ and\ \bibinfo {author} {\bibfnamefont
  {G.}~\bibnamefont {Leus}},\ }\bibfield  {title} {\bibinfo {title}
  {Compressive 3d ultrasound imaging using a single sensor},\ }\bibfield
  {journal} {\bibinfo  {journal} {Science Advances}\ }\textbf {\bibinfo
  {volume} {3}},\ \href {https://doi.org/10.1126/sciadv.1701423}
  {10.1126/sciadv.1701423} (\bibinfo {year} {2017})\BibitemShut {NoStop}%
\bibitem [{\citenamefont {Antipa}\ \emph {et~al.}(2018)\citenamefont {Antipa},
  \citenamefont {Kuo}, \citenamefont {Heckel}, \citenamefont {Mildenhall},
  \citenamefont {Bostan}, \citenamefont {Ng},\ and\ \citenamefont
  {Waller}}]{Antipa2018}%
  \BibitemOpen
  \bibfield  {author} {\bibinfo {author} {\bibfnamefont {N.}~\bibnamefont
  {Antipa}}, \bibinfo {author} {\bibfnamefont {G.}~\bibnamefont {Kuo}},
  \bibinfo {author} {\bibfnamefont {R.}~\bibnamefont {Heckel}}, \bibinfo
  {author} {\bibfnamefont {B.}~\bibnamefont {Mildenhall}}, \bibinfo {author}
  {\bibfnamefont {E.}~\bibnamefont {Bostan}}, \bibinfo {author} {\bibfnamefont
  {R.}~\bibnamefont {Ng}},\ and\ \bibinfo {author} {\bibfnamefont
  {L.}~\bibnamefont {Waller}},\ }\bibfield  {title} {\bibinfo {title}
  {Diffusercam: lensless single-exposure 3d imaging},\ }\href@noop {}
  {\bibfield  {journal} {\bibinfo  {journal} {Optica}\ }\textbf {\bibinfo
  {volume} {5}},\ \bibinfo {pages} {1} (\bibinfo {year} {2018})}\BibitemShut
  {NoStop}%
\bibitem [{\citenamefont {Malone}\ \emph {et~al.}(2023)\citenamefont {Malone},
  \citenamefont {Aggarwal}, \citenamefont {Waller},\ and\ \citenamefont
  {Bowden}}]{Malone2023}%
  \BibitemOpen
  \bibfield  {author} {\bibinfo {author} {\bibfnamefont {J.~D.}\ \bibnamefont
  {Malone}}, \bibinfo {author} {\bibfnamefont {N.}~\bibnamefont {Aggarwal}},
  \bibinfo {author} {\bibfnamefont {L.}~\bibnamefont {Waller}},\ and\ \bibinfo
  {author} {\bibfnamefont {A.~K.}\ \bibnamefont {Bowden}},\ }\bibfield  {title}
  {\bibinfo {title} {Diffuserspec: spectroscopy with scotch tape},\ }\href@noop
  {} {\bibfield  {journal} {\bibinfo  {journal} {Optics Letters}\ }\textbf
  {\bibinfo {volume} {48}},\ \bibinfo {pages} {323} (\bibinfo {year}
  {2023})}\BibitemShut {NoStop}%
\bibitem [{\citenamefont {Tondo~Yoya}\ \emph {et~al.}(2017)\citenamefont
  {Tondo~Yoya}, \citenamefont {Fuchs},\ and\ \citenamefont {Davy}}]{Tondo2017}%
  \BibitemOpen
  \bibfield  {author} {\bibinfo {author} {\bibfnamefont {A.~C.}\ \bibnamefont
  {Tondo~Yoya}}, \bibinfo {author} {\bibfnamefont {B.}~\bibnamefont {Fuchs}},\
  and\ \bibinfo {author} {\bibfnamefont {M.}~\bibnamefont {Davy}},\ }\bibfield
  {title} {\bibinfo {title} {Computational passive imaging of thermal sources
  with a leaky chaotic cavity},\ }\href@noop {} {\bibfield  {journal} {\bibinfo
   {journal} {Applied Physics Letters}\ }\textbf {\bibinfo {volume} {111}},\
  \bibinfo {pages} {193501} (\bibinfo {year} {2017})}\BibitemShut {NoStop}%
\bibitem [{\citenamefont {Fromenteze}\ \emph {et~al.}(2015)\citenamefont
  {Fromenteze}, \citenamefont {Yurduseven}, \citenamefont {Imani},
  \citenamefont {Gollub}, \citenamefont {Decroze}, \citenamefont {Carsenat},\
  and\ \citenamefont {Smith}}]{Fromenteze2015}%
  \BibitemOpen
  \bibfield  {author} {\bibinfo {author} {\bibfnamefont {T.}~\bibnamefont
  {Fromenteze}}, \bibinfo {author} {\bibfnamefont {O.}~\bibnamefont
  {Yurduseven}}, \bibinfo {author} {\bibfnamefont {M.~F.}\ \bibnamefont
  {Imani}}, \bibinfo {author} {\bibfnamefont {J.}~\bibnamefont {Gollub}},
  \bibinfo {author} {\bibfnamefont {C.}~\bibnamefont {Decroze}}, \bibinfo
  {author} {\bibfnamefont {D.}~\bibnamefont {Carsenat}},\ and\ \bibinfo
  {author} {\bibfnamefont {D.~R.}\ \bibnamefont {Smith}},\ }\bibfield  {title}
  {\bibinfo {title} {Computational imaging using a mode-mixing cavity at
  microwave frequencies},\ }\href@noop {} {\bibfield  {journal} {\bibinfo
  {journal} {Applied Physics Letters}\ }\textbf {\bibinfo {volume} {106}},\
  \bibinfo {pages} {194104} (\bibinfo {year} {2015})}\BibitemShut {NoStop}%
\bibitem [{\citenamefont {Edgar}\ \emph {et~al.}(2019)\citenamefont {Edgar},
  \citenamefont {Gibson},\ and\ \citenamefont {Padgett}}]{edgar2019principles}%
  \BibitemOpen
  \bibfield  {author} {\bibinfo {author} {\bibfnamefont {M.~P.}\ \bibnamefont
  {Edgar}}, \bibinfo {author} {\bibfnamefont {G.~M.}\ \bibnamefont {Gibson}},\
  and\ \bibinfo {author} {\bibfnamefont {M.~J.}\ \bibnamefont {Padgett}},\
  }\bibfield  {title} {\bibinfo {title} {Principles and prospects for
  single-pixel imaging},\ }\href@noop {} {\bibfield  {journal} {\bibinfo
  {journal} {Nature photonics}\ }\textbf {\bibinfo {volume} {13}},\ \bibinfo
  {pages} {13} (\bibinfo {year} {2019})}\BibitemShut {NoStop}%
\bibitem [{\citenamefont {Mait}\ \emph {et~al.}(2018)\citenamefont {Mait},
  \citenamefont {Euliss},\ and\ \citenamefont
  {Athale}}]{mait2018computational}%
  \BibitemOpen
  \bibfield  {author} {\bibinfo {author} {\bibfnamefont {J.~N.}\ \bibnamefont
  {Mait}}, \bibinfo {author} {\bibfnamefont {G.~W.}\ \bibnamefont {Euliss}},\
  and\ \bibinfo {author} {\bibfnamefont {R.~A.}\ \bibnamefont {Athale}},\
  }\bibfield  {title} {\bibinfo {title} {Computational imaging},\ }\href@noop
  {} {\bibfield  {journal} {\bibinfo  {journal} {Advances in Optics and
  Photonics}\ }\textbf {\bibinfo {volume} {10}},\ \bibinfo {pages} {409}
  (\bibinfo {year} {2018})}\BibitemShut {NoStop}%
\bibitem [{\citenamefont {Draeger}\ and\ \citenamefont
  {Fink}(1999)}]{draeger1999one2}%
  \BibitemOpen
  \bibfield  {author} {\bibinfo {author} {\bibfnamefont {C.}~\bibnamefont
  {Draeger}}\ and\ \bibinfo {author} {\bibfnamefont {M.}~\bibnamefont {Fink}},\
  }\bibfield  {title} {\bibinfo {title} {One-channel time-reversal in chaotic
  cavities: Theoretical limits},\ }\href@noop {} {\bibfield  {journal}
  {\bibinfo  {journal} {The Journal of the Acoustical Society of America}\
  }\textbf {\bibinfo {volume} {105}},\ \bibinfo {pages} {611} (\bibinfo {year}
  {1999})}\BibitemShut {NoStop}%
\bibitem [{\citenamefont {Derode}\ \emph {et~al.}(1999)\citenamefont {Derode},
  \citenamefont {Tourin},\ and\ \citenamefont {Fink}}]{derode1999ultrasonic}%
  \BibitemOpen
  \bibfield  {author} {\bibinfo {author} {\bibfnamefont {A.}~\bibnamefont
  {Derode}}, \bibinfo {author} {\bibfnamefont {A.}~\bibnamefont {Tourin}},\
  and\ \bibinfo {author} {\bibfnamefont {M.}~\bibnamefont {Fink}},\ }\bibfield
  {title} {\bibinfo {title} {Ultrasonic pulse compression with one-bit time
  reversal through multiple scattering},\ }\href@noop {} {\bibfield  {journal}
  {\bibinfo  {journal} {Journal of applied physics}\ }\textbf {\bibinfo
  {volume} {85}},\ \bibinfo {pages} {6343} (\bibinfo {year}
  {1999})}\BibitemShut {NoStop}%
\bibitem [{\citenamefont {Montaldo}\ \emph {et~al.}(2004)\citenamefont
  {Montaldo}, \citenamefont {Palacio}, \citenamefont {Tanter},\ and\
  \citenamefont {Fink}}]{Montaldo2004}%
  \BibitemOpen
  \bibfield  {author} {\bibinfo {author} {\bibfnamefont {G.}~\bibnamefont
  {Montaldo}}, \bibinfo {author} {\bibfnamefont {D.}~\bibnamefont {Palacio}},
  \bibinfo {author} {\bibfnamefont {M.}~\bibnamefont {Tanter}},\ and\ \bibinfo
  {author} {\bibfnamefont {M.}~\bibnamefont {Fink}},\ }\bibfield  {title}
  {\bibinfo {title} {Time reversal kaleidoscope: A smart transducer for
  three-dimensional ultrasonic imaging},\ }\href@noop {} {\bibfield  {journal}
  {\bibinfo  {journal} {Applied physics letters}\ }\textbf {\bibinfo {volume}
  {84}},\ \bibinfo {pages} {3879} (\bibinfo {year} {2004})}\BibitemShut
  {NoStop}%
\bibitem [{\citenamefont {Robin}\ \emph {et~al.}(2018)\citenamefont {Robin},
  \citenamefont {Simon}, \citenamefont {Arnal}, \citenamefont {Tanter},\ and\
  \citenamefont {Pernot}}]{Robin2018}%
  \BibitemOpen
  \bibfield  {author} {\bibinfo {author} {\bibfnamefont {J.}~\bibnamefont
  {Robin}}, \bibinfo {author} {\bibfnamefont {A.}~\bibnamefont {Simon}},
  \bibinfo {author} {\bibfnamefont {B.}~\bibnamefont {Arnal}}, \bibinfo
  {author} {\bibfnamefont {M.}~\bibnamefont {Tanter}},\ and\ \bibinfo {author}
  {\bibfnamefont {M.}~\bibnamefont {Pernot}},\ }\bibfield  {title} {\bibinfo
  {title} {Self-adaptive ultrasonic beam amplifiers: Application to transcostal
  shock wave therapy},\ }\href@noop {} {\bibfield  {journal} {\bibinfo
  {journal} {Physics in Medicine \& Biology}\ }\textbf {\bibinfo {volume}
  {63}},\ \bibinfo {pages} {175014} (\bibinfo {year} {2018})}\BibitemShut
  {NoStop}%
\bibitem [{\citenamefont {Caron-Grenier}\ \emph {et~al.}(2023)\citenamefont
  {Caron-Grenier}, \citenamefont {Poree}, \citenamefont {Perrot}, \citenamefont
  {Ramos-Palacio}, \citenamefont {Sadikot},\ and\ \citenamefont
  {Provost}}]{caron2023ergodic}%
  \BibitemOpen
  \bibfield  {author} {\bibinfo {author} {\bibfnamefont {O.}~\bibnamefont
  {Caron-Grenier}}, \bibinfo {author} {\bibfnamefont {J.}~\bibnamefont
  {Poree}}, \bibinfo {author} {\bibfnamefont {V.}~\bibnamefont {Perrot}},
  \bibinfo {author} {\bibfnamefont {G.}~\bibnamefont {Ramos-Palacio}}, \bibinfo
  {author} {\bibfnamefont {A.~F.}\ \bibnamefont {Sadikot}},\ and\ \bibinfo
  {author} {\bibfnamefont {J.}~\bibnamefont {Provost}},\ }\bibfield  {title}
  {\bibinfo {title} {Ergodic encoding for single-element ultrasound imaging in
  vivo},\ }\href@noop {} {\bibfield  {journal} {\bibinfo  {journal} {arXiv
  preprint arXiv:2308.09196}\ } (\bibinfo {year} {2023})}\BibitemShut {NoStop}%
\bibitem [{\citenamefont {Li}\ \emph {et~al.}(2020{\natexlab{a}})\citenamefont
  {Li}, \citenamefont {Li}, \citenamefont {Zhu}, \citenamefont {Maslov},
  \citenamefont {Shi}, \citenamefont {Hu}, \citenamefont {Bo}, \citenamefont
  {Yao}, \citenamefont {Liang}, \citenamefont {Wang} \emph {et~al.}}]{Li2020}%
  \BibitemOpen
  \bibfield  {author} {\bibinfo {author} {\bibfnamefont {Y.}~\bibnamefont
  {Li}}, \bibinfo {author} {\bibfnamefont {L.}~\bibnamefont {Li}}, \bibinfo
  {author} {\bibfnamefont {L.}~\bibnamefont {Zhu}}, \bibinfo {author}
  {\bibfnamefont {K.}~\bibnamefont {Maslov}}, \bibinfo {author} {\bibfnamefont
  {J.}~\bibnamefont {Shi}}, \bibinfo {author} {\bibfnamefont {P.}~\bibnamefont
  {Hu}}, \bibinfo {author} {\bibfnamefont {E.}~\bibnamefont {Bo}}, \bibinfo
  {author} {\bibfnamefont {J.}~\bibnamefont {Yao}}, \bibinfo {author}
  {\bibfnamefont {J.}~\bibnamefont {Liang}}, \bibinfo {author} {\bibfnamefont
  {L.}~\bibnamefont {Wang}}, \emph {et~al.},\ }\bibfield  {title} {\bibinfo
  {title} {Snapshot photoacoustic topography through an ergodic relay for
  high-throughput imaging of optical absorption},\ }\href@noop {} {\bibfield
  {journal} {\bibinfo  {journal} {Nature Photonics}\ }\textbf {\bibinfo
  {volume} {14}},\ \bibinfo {pages} {164} (\bibinfo {year}
  {2020}{\natexlab{a}})}\BibitemShut {NoStop}%
\bibitem [{\citenamefont {Li}\ \emph {et~al.}(2020{\natexlab{b}})\citenamefont
  {Li}, \citenamefont {Wong}, \citenamefont {Shi}, \citenamefont {Hsu},\ and\
  \citenamefont {Wang}}]{Li20202}%
  \BibitemOpen
  \bibfield  {author} {\bibinfo {author} {\bibfnamefont {Y.}~\bibnamefont
  {Li}}, \bibinfo {author} {\bibfnamefont {T.~T.}\ \bibnamefont {Wong}},
  \bibinfo {author} {\bibfnamefont {J.}~\bibnamefont {Shi}}, \bibinfo {author}
  {\bibfnamefont {H.-C.}\ \bibnamefont {Hsu}},\ and\ \bibinfo {author}
  {\bibfnamefont {L.~V.}\ \bibnamefont {Wang}},\ }\bibfield  {title} {\bibinfo
  {title} {Multifocal photoacoustic microscopy using a single-element
  ultrasonic transducer through an ergodic relay},\ }\href@noop {} {\bibfield
  {journal} {\bibinfo  {journal} {Light: Science \& Applications}\ }\textbf
  {\bibinfo {volume} {9}},\ \bibinfo {pages} {135} (\bibinfo {year}
  {2020}{\natexlab{b}})}\BibitemShut {NoStop}%
\bibitem [{\citenamefont {Li}\ \emph {et~al.}(2021)\citenamefont {Li},
  \citenamefont {Li}, \citenamefont {Zhang},\ and\ \citenamefont
  {Wang}}]{Li2021}%
  \BibitemOpen
  \bibfield  {author} {\bibinfo {author} {\bibfnamefont {L.}~\bibnamefont
  {Li}}, \bibinfo {author} {\bibfnamefont {Y.}~\bibnamefont {Li}}, \bibinfo
  {author} {\bibfnamefont {Y.}~\bibnamefont {Zhang}},\ and\ \bibinfo {author}
  {\bibfnamefont {L.~V.}\ \bibnamefont {Wang}},\ }\bibfield  {title} {\bibinfo
  {title} {Snapshot photoacoustic topography through an ergodic relay of
  optical absorption in vivo},\ }\href@noop {} {\bibfield  {journal} {\bibinfo
  {journal} {Nature protocols}\ }\textbf {\bibinfo {volume} {16}},\ \bibinfo
  {pages} {2381} (\bibinfo {year} {2021})}\BibitemShut {NoStop}%
\bibitem [{\citenamefont {Janjic}\ \emph
  {et~al.}(2018{\natexlab{b}})\citenamefont {Janjic}, \citenamefont
  {Kruizinga}, \citenamefont {Van Der~Meulen}, \citenamefont {Springeling},
  \citenamefont {Mastik}, \citenamefont {Leus}, \citenamefont {Bosch},
  \citenamefont {van~der Steen},\ and\ \citenamefont {van
  Soest}}]{janjic2018structured}%
  \BibitemOpen
  \bibfield  {author} {\bibinfo {author} {\bibfnamefont {J.}~\bibnamefont
  {Janjic}}, \bibinfo {author} {\bibfnamefont {P.}~\bibnamefont {Kruizinga}},
  \bibinfo {author} {\bibfnamefont {P.}~\bibnamefont {Van Der~Meulen}},
  \bibinfo {author} {\bibfnamefont {G.}~\bibnamefont {Springeling}}, \bibinfo
  {author} {\bibfnamefont {F.}~\bibnamefont {Mastik}}, \bibinfo {author}
  {\bibfnamefont {G.}~\bibnamefont {Leus}}, \bibinfo {author} {\bibfnamefont
  {J.~G.}\ \bibnamefont {Bosch}}, \bibinfo {author} {\bibfnamefont {A.~F.}\
  \bibnamefont {van~der Steen}},\ and\ \bibinfo {author} {\bibfnamefont
  {G.}~\bibnamefont {van Soest}},\ }\bibfield  {title} {\bibinfo {title}
  {Structured ultrasound microscopy},\ }\href@noop {} {\bibfield  {journal}
  {\bibinfo  {journal} {Applied Physics Letters}\ }\textbf {\bibinfo {volume}
  {112}} (\bibinfo {year} {2018}{\natexlab{b}})}\BibitemShut {NoStop}%
\bibitem [{\citenamefont {Jensen}\ \emph {et~al.}(2006)\citenamefont {Jensen},
  \citenamefont {Nikolov}, \citenamefont {Gammelmark},\ and\ \citenamefont
  {Pedersen}}]{Jensen2006}%
  \BibitemOpen
  \bibfield  {author} {\bibinfo {author} {\bibfnamefont {J.~A.}\ \bibnamefont
  {Jensen}}, \bibinfo {author} {\bibfnamefont {S.~I.}\ \bibnamefont {Nikolov}},
  \bibinfo {author} {\bibfnamefont {K.~L.}\ \bibnamefont {Gammelmark}},\ and\
  \bibinfo {author} {\bibfnamefont {M.~H.}\ \bibnamefont {Pedersen}},\
  }\bibfield  {title} {\bibinfo {title} {Synthetic aperture ultrasound
  imaging},\ }\href@noop {} {\bibfield  {journal} {\bibinfo  {journal}
  {Ultrasonics}\ }\textbf {\bibinfo {volume} {44}},\ \bibinfo {pages} {e5}
  (\bibinfo {year} {2006})}\BibitemShut {NoStop}%
\bibitem [{\citenamefont {Chiao}\ \emph {et~al.}(1997)\citenamefont {Chiao},
  \citenamefont {Thomas},\ and\ \citenamefont {Silverstein}}]{Chiao1997}%
  \BibitemOpen
  \bibfield  {author} {\bibinfo {author} {\bibfnamefont {R.~Y.}\ \bibnamefont
  {Chiao}}, \bibinfo {author} {\bibfnamefont {L.~J.}\ \bibnamefont {Thomas}},\
  and\ \bibinfo {author} {\bibfnamefont {S.~D.}\ \bibnamefont {Silverstein}},\
  }\bibfield  {title} {\bibinfo {title} {Sparse array imaging with
  spatially-encoded transmits},\ }in\ \href@noop {} {\emph {\bibinfo
  {booktitle} {1997 IEEE Ultrasonics Symposium Proceedings. An International
  Symposium (Cat. No. 97CH36118)}}},\ Vol.~\bibinfo {volume} {2}\ (\bibinfo
  {organization} {IEEE},\ \bibinfo {year} {1997})\ pp.\ \bibinfo {pages}
  {1679--1682}\BibitemShut {NoStop}%
\bibitem [{\citenamefont {Demen{\'e}}\ \emph {et~al.}(2015)\citenamefont
  {Demen{\'e}}, \citenamefont {Deffieux}, \citenamefont {Pernot}, \citenamefont
  {Osmanski}, \citenamefont {Biran}, \citenamefont {Gennisson}, \citenamefont
  {Sieu}, \citenamefont {Bergel}, \citenamefont {Franqui}, \citenamefont
  {Correas} \emph {et~al.}}]{demene2015spatiotemporal}%
  \BibitemOpen
  \bibfield  {author} {\bibinfo {author} {\bibfnamefont {C.}~\bibnamefont
  {Demen{\'e}}}, \bibinfo {author} {\bibfnamefont {T.}~\bibnamefont
  {Deffieux}}, \bibinfo {author} {\bibfnamefont {M.}~\bibnamefont {Pernot}},
  \bibinfo {author} {\bibfnamefont {B.-F.}\ \bibnamefont {Osmanski}}, \bibinfo
  {author} {\bibfnamefont {V.}~\bibnamefont {Biran}}, \bibinfo {author}
  {\bibfnamefont {J.-L.}\ \bibnamefont {Gennisson}}, \bibinfo {author}
  {\bibfnamefont {L.-A.}\ \bibnamefont {Sieu}}, \bibinfo {author}
  {\bibfnamefont {A.}~\bibnamefont {Bergel}}, \bibinfo {author} {\bibfnamefont
  {S.}~\bibnamefont {Franqui}}, \bibinfo {author} {\bibfnamefont {J.-M.}\
  \bibnamefont {Correas}}, \emph {et~al.},\ }\bibfield  {title} {\bibinfo
  {title} {Spatiotemporal clutter filtering of ultrafast ultrasound data highly
  increases doppler and fultrasound sensitivity},\ }\href@noop {} {\bibfield
  {journal} {\bibinfo  {journal} {IEEE transactions on medical imaging}\
  }\textbf {\bibinfo {volume} {34}},\ \bibinfo {pages} {2271} (\bibinfo {year}
  {2015})}\BibitemShut {NoStop}%
\bibitem [{\citenamefont {Madiena}\ \emph {et~al.}(2018)\citenamefont
  {Madiena}, \citenamefont {Faurie}, \citenamefont {Por{\'e}e},\ and\
  \citenamefont {Garcia}}]{madiena2018color}%
  \BibitemOpen
  \bibfield  {author} {\bibinfo {author} {\bibfnamefont {C.}~\bibnamefont
  {Madiena}}, \bibinfo {author} {\bibfnamefont {J.}~\bibnamefont {Faurie}},
  \bibinfo {author} {\bibfnamefont {J.}~\bibnamefont {Por{\'e}e}},\ and\
  \bibinfo {author} {\bibfnamefont {D.}~\bibnamefont {Garcia}},\ }\bibfield
  {title} {\bibinfo {title} {Color and vector flow imaging in parallel
  ultrasound with sub-nyquist sampling},\ }\href@noop {} {\bibfield  {journal}
  {\bibinfo  {journal} {IEEE transactions on ultrasonics, ferroelectrics, and
  frequency control}\ }\textbf {\bibinfo {volume} {65}},\ \bibinfo {pages}
  {795} (\bibinfo {year} {2018})}\BibitemShut {NoStop}%
\bibitem [{\citenamefont {Hu}\ \emph {et~al.}(2023)\citenamefont {Hu},
  \citenamefont {Huang}, \citenamefont {Li}, \citenamefont {Gao}, \citenamefont
  {Yin}, \citenamefont {Qi}, \citenamefont {Wu}, \citenamefont {Chen},
  \citenamefont {Ma}, \citenamefont {Shi} \emph {et~al.}}]{hu2023wearable}%
  \BibitemOpen
  \bibfield  {author} {\bibinfo {author} {\bibfnamefont {H.}~\bibnamefont
  {Hu}}, \bibinfo {author} {\bibfnamefont {H.}~\bibnamefont {Huang}}, \bibinfo
  {author} {\bibfnamefont {M.}~\bibnamefont {Li}}, \bibinfo {author}
  {\bibfnamefont {X.}~\bibnamefont {Gao}}, \bibinfo {author} {\bibfnamefont
  {L.}~\bibnamefont {Yin}}, \bibinfo {author} {\bibfnamefont {R.}~\bibnamefont
  {Qi}}, \bibinfo {author} {\bibfnamefont {R.~S.}\ \bibnamefont {Wu}}, \bibinfo
  {author} {\bibfnamefont {X.}~\bibnamefont {Chen}}, \bibinfo {author}
  {\bibfnamefont {Y.}~\bibnamefont {Ma}}, \bibinfo {author} {\bibfnamefont
  {K.}~\bibnamefont {Shi}}, \emph {et~al.},\ }\bibfield  {title} {\bibinfo
  {title} {A wearable cardiac ultrasound imager},\ }\href@noop {} {\bibfield
  {journal} {\bibinfo  {journal} {Nature}\ }\textbf {\bibinfo {volume} {613}},\
  \bibinfo {pages} {667} (\bibinfo {year} {2023})}\BibitemShut {NoStop}%
\bibitem [{\citenamefont {Demen{\'e}}\ \emph {et~al.}(2021)\citenamefont
  {Demen{\'e}}, \citenamefont {Robin}, \citenamefont {Dizeux}, \citenamefont
  {Heiles}, \citenamefont {Pernot}, \citenamefont {Tanter},\ and\ \citenamefont
  {Perren}}]{demene2021transcranial}%
  \BibitemOpen
  \bibfield  {author} {\bibinfo {author} {\bibfnamefont {C.}~\bibnamefont
  {Demen{\'e}}}, \bibinfo {author} {\bibfnamefont {J.}~\bibnamefont {Robin}},
  \bibinfo {author} {\bibfnamefont {A.}~\bibnamefont {Dizeux}}, \bibinfo
  {author} {\bibfnamefont {B.}~\bibnamefont {Heiles}}, \bibinfo {author}
  {\bibfnamefont {M.}~\bibnamefont {Pernot}}, \bibinfo {author} {\bibfnamefont
  {M.}~\bibnamefont {Tanter}},\ and\ \bibinfo {author} {\bibfnamefont
  {F.}~\bibnamefont {Perren}},\ }\bibfield  {title} {\bibinfo {title}
  {Transcranial ultrafast ultrasound localization microscopy of brain
  vasculature in patients},\ }\href@noop {} {\bibfield  {journal} {\bibinfo
  {journal} {Nature biomedical engineering}\ }\textbf {\bibinfo {volume} {5}},\
  \bibinfo {pages} {219} (\bibinfo {year} {2021})}\BibitemShut {NoStop}%
\bibitem [{\citenamefont {Molesky}\ \emph {et~al.}(2018)\citenamefont
  {Molesky}, \citenamefont {Lin}, \citenamefont {Piggott}, \citenamefont {Jin},
  \citenamefont {Vuckovi{\'c}},\ and\ \citenamefont
  {Rodriguez}}]{molesky2018inverse}%
  \BibitemOpen
  \bibfield  {author} {\bibinfo {author} {\bibfnamefont {S.}~\bibnamefont
  {Molesky}}, \bibinfo {author} {\bibfnamefont {Z.}~\bibnamefont {Lin}},
  \bibinfo {author} {\bibfnamefont {A.~Y.}\ \bibnamefont {Piggott}}, \bibinfo
  {author} {\bibfnamefont {W.}~\bibnamefont {Jin}}, \bibinfo {author}
  {\bibfnamefont {J.}~\bibnamefont {Vuckovi{\'c}}},\ and\ \bibinfo {author}
  {\bibfnamefont {A.~W.}\ \bibnamefont {Rodriguez}},\ }\bibfield  {title}
  {\bibinfo {title} {Inverse design in nanophotonics},\ }\href@noop {}
  {\bibfield  {journal} {\bibinfo  {journal} {Nature Photonics}\ }\textbf
  {\bibinfo {volume} {12}},\ \bibinfo {pages} {659} (\bibinfo {year}
  {2018})}\BibitemShut {NoStop}%
\bibitem [{\citenamefont {Kellman}\ \emph {et~al.}(2019)\citenamefont
  {Kellman}, \citenamefont {Bostan}, \citenamefont {Repina},\ and\
  \citenamefont {Waller}}]{kellman2019physics}%
  \BibitemOpen
  \bibfield  {author} {\bibinfo {author} {\bibfnamefont {M.~R.}\ \bibnamefont
  {Kellman}}, \bibinfo {author} {\bibfnamefont {E.}~\bibnamefont {Bostan}},
  \bibinfo {author} {\bibfnamefont {N.~A.}\ \bibnamefont {Repina}},\ and\
  \bibinfo {author} {\bibfnamefont {L.}~\bibnamefont {Waller}},\ }\bibfield
  {title} {\bibinfo {title} {Physics-based learned design: optimized
  coded-illumination for quantitative phase imaging},\ }\href@noop {}
  {\bibfield  {journal} {\bibinfo  {journal} {IEEE Transactions on
  Computational Imaging}\ }\textbf {\bibinfo {volume} {5}},\ \bibinfo {pages}
  {344} (\bibinfo {year} {2019})}\BibitemShut {NoStop}%
\bibitem [{\citenamefont {Ongie}\ \emph {et~al.}(2020)\citenamefont {Ongie},
  \citenamefont {Jalal}, \citenamefont {Metzler}, \citenamefont {Baraniuk},
  \citenamefont {Dimakis},\ and\ \citenamefont {Willett}}]{ongie2020deep}%
  \BibitemOpen
  \bibfield  {author} {\bibinfo {author} {\bibfnamefont {G.}~\bibnamefont
  {Ongie}}, \bibinfo {author} {\bibfnamefont {A.}~\bibnamefont {Jalal}},
  \bibinfo {author} {\bibfnamefont {C.~A.}\ \bibnamefont {Metzler}}, \bibinfo
  {author} {\bibfnamefont {R.~G.}\ \bibnamefont {Baraniuk}}, \bibinfo {author}
  {\bibfnamefont {A.~G.}\ \bibnamefont {Dimakis}},\ and\ \bibinfo {author}
  {\bibfnamefont {R.}~\bibnamefont {Willett}},\ }\bibfield  {title} {\bibinfo
  {title} {Deep learning techniques for inverse problems in imaging},\
  }\href@noop {} {\bibfield  {journal} {\bibinfo  {journal} {IEEE Journal on
  Selected Areas in Information Theory}\ }\textbf {\bibinfo {volume} {1}},\
  \bibinfo {pages} {39} (\bibinfo {year} {2020})}\BibitemShut {NoStop}%
\bibitem [{\citenamefont {Melde}\ \emph {et~al.}(2016)\citenamefont {Melde},
  \citenamefont {Mark}, \citenamefont {Qiu},\ and\ \citenamefont
  {Fischer}}]{Melde2016}%
  \BibitemOpen
  \bibfield  {author} {\bibinfo {author} {\bibfnamefont {K.}~\bibnamefont
  {Melde}}, \bibinfo {author} {\bibfnamefont {A.~G.}\ \bibnamefont {Mark}},
  \bibinfo {author} {\bibfnamefont {T.}~\bibnamefont {Qiu}},\ and\ \bibinfo
  {author} {\bibfnamefont {P.}~\bibnamefont {Fischer}},\ }\bibfield  {title}
  {\bibinfo {title} {Holograms for acoustics},\ }\href
  {https://doi.org/10.1038/nature19755} {\bibfield  {journal} {\bibinfo
  {journal} {Nature}\ }\textbf {\bibinfo {volume} {537}},\ \bibinfo {pages}
  {518} (\bibinfo {year} {2016})}\BibitemShut {NoStop}%
\bibitem [{\citenamefont {Lopes}\ \emph {et~al.}(2017)\citenamefont {Lopes},
  \citenamefont {Andrade}, \citenamefont {Leao-Neto}, \citenamefont
  {Adamowski}, \citenamefont {Minin},\ and\ \citenamefont {Silva}}]{Lopes2017}%
  \BibitemOpen
  \bibfield  {author} {\bibinfo {author} {\bibfnamefont {J.}~\bibnamefont
  {Lopes}}, \bibinfo {author} {\bibfnamefont {M.}~\bibnamefont {Andrade}},
  \bibinfo {author} {\bibfnamefont {J.}~\bibnamefont {Leao-Neto}}, \bibinfo
  {author} {\bibfnamefont {J.}~\bibnamefont {Adamowski}}, \bibinfo {author}
  {\bibfnamefont {I.}~\bibnamefont {Minin}},\ and\ \bibinfo {author}
  {\bibfnamefont {G.}~\bibnamefont {Silva}},\ }\bibfield  {title} {\bibinfo
  {title} {Focusing acoustic beams with a ball-shaped lens beyond the
  diffraction limit},\ }\href@noop {} {\bibfield  {journal} {\bibinfo
  {journal} {Physical Review Applied}\ }\textbf {\bibinfo {volume} {8}},\
  \bibinfo {pages} {024013} (\bibinfo {year} {2017})}\BibitemShut {NoStop}%
\bibitem [{\citenamefont {Markley}\ \emph {et~al.}(2021)\citenamefont
  {Markley}, \citenamefont {Liu}, \citenamefont {Kellman}, \citenamefont
  {Antipa},\ and\ \citenamefont {Waller}}]{markley2021physics}%
  \BibitemOpen
  \bibfield  {author} {\bibinfo {author} {\bibfnamefont {E.}~\bibnamefont
  {Markley}}, \bibinfo {author} {\bibfnamefont {F.~L.}\ \bibnamefont {Liu}},
  \bibinfo {author} {\bibfnamefont {M.}~\bibnamefont {Kellman}}, \bibinfo
  {author} {\bibfnamefont {N.}~\bibnamefont {Antipa}},\ and\ \bibinfo {author}
  {\bibfnamefont {L.}~\bibnamefont {Waller}},\ }\bibfield  {title} {\bibinfo
  {title} {Physics-based learned diffuser for single-shot 3d imaging},\ }in\
  \href@noop {} {\emph {\bibinfo {booktitle} {NeurIPS 2021 Workshop on Deep
  Learning and Inverse Problems}}}\ (\bibinfo {year} {2021})\BibitemShut
  {NoStop}%
\bibitem [{\citenamefont {Holmes}\ \emph {et~al.}(2005)\citenamefont {Holmes},
  \citenamefont {Drinkwater},\ and\ \citenamefont {Wilcox}}]{Holmes2005}%
  \BibitemOpen
  \bibfield  {author} {\bibinfo {author} {\bibfnamefont {C.}~\bibnamefont
  {Holmes}}, \bibinfo {author} {\bibfnamefont {B.~W.}\ \bibnamefont
  {Drinkwater}},\ and\ \bibinfo {author} {\bibfnamefont {P.~D.}\ \bibnamefont
  {Wilcox}},\ }\bibfield  {title} {\bibinfo {title} {Post-processing of the
  full matrix of ultrasonic transmit--receive array data for non-destructive
  evaluation},\ }\href@noop {} {\bibfield  {journal} {\bibinfo  {journal} {NDT
  \& e International}\ }\textbf {\bibinfo {volume} {38}},\ \bibinfo {pages}
  {701} (\bibinfo {year} {2005})}\BibitemShut {NoStop}%
\bibitem [{\citenamefont {Bertolo}\ \emph {et~al.}(2021)\citenamefont
  {Bertolo}, \citenamefont {Nouhoum}, \citenamefont {Cazzanelli}, \citenamefont
  {Ferrier}, \citenamefont {Mariani}, \citenamefont {Kliewer}, \citenamefont
  {Belliard}, \citenamefont {Osmanski}, \citenamefont {Deffieux}, \citenamefont
  {Pezet} \emph {et~al.}}]{bertolo2021whole}%
  \BibitemOpen
  \bibfield  {author} {\bibinfo {author} {\bibfnamefont {A.}~\bibnamefont
  {Bertolo}}, \bibinfo {author} {\bibfnamefont {M.}~\bibnamefont {Nouhoum}},
  \bibinfo {author} {\bibfnamefont {S.}~\bibnamefont {Cazzanelli}}, \bibinfo
  {author} {\bibfnamefont {J.}~\bibnamefont {Ferrier}}, \bibinfo {author}
  {\bibfnamefont {J.-C.}\ \bibnamefont {Mariani}}, \bibinfo {author}
  {\bibfnamefont {A.}~\bibnamefont {Kliewer}}, \bibinfo {author} {\bibfnamefont
  {B.}~\bibnamefont {Belliard}}, \bibinfo {author} {\bibfnamefont {B.-F.}\
  \bibnamefont {Osmanski}}, \bibinfo {author} {\bibfnamefont {T.}~\bibnamefont
  {Deffieux}}, \bibinfo {author} {\bibfnamefont {S.}~\bibnamefont {Pezet}},
  \emph {et~al.},\ }\bibfield  {title} {\bibinfo {title} {Whole-brain 3d
  activation and functional connectivity mapping in mice using transcranial
  functional ultrasound imaging},\ }\href@noop {} {\bibfield  {journal}
  {\bibinfo  {journal} {Journal of visualized experiments: JoVE}\ ,\ \bibinfo
  {pages} {e62267}} (\bibinfo {year} {2021})}\BibitemShut {NoStop}%
\bibitem [{\citenamefont {Kaczkowski}(2016)}]{kaczkowski2016bandwidth}%
  \BibitemOpen
  \bibfield  {author} {\bibinfo {author} {\bibfnamefont {P.}~\bibnamefont
  {Kaczkowski}},\ }\bibfield  {title} {\bibinfo {title} {Bandwidth sampling
  data acquisition with the vantage system for high frequency transducers},\
  }\href@noop {} {\bibfield  {journal} {\bibinfo  {journal} {Verasonics White
  Pap}\ ,\ \bibinfo {pages} {1}} (\bibinfo {year} {2016})}\BibitemShut
  {NoStop}%
\bibitem [{\citenamefont {Bae}\ \emph {et~al.}(2002)\citenamefont {Bae},
  \citenamefont {Lee}, \citenamefont {Jeong},\ and\ \citenamefont
  {Kwon}}]{bae2002orthogonal}%
  \BibitemOpen
  \bibfield  {author} {\bibinfo {author} {\bibfnamefont {M.-H.}\ \bibnamefont
  {Bae}}, \bibinfo {author} {\bibfnamefont {W.-Y.}\ \bibnamefont {Lee}},
  \bibinfo {author} {\bibfnamefont {M.-K.}\ \bibnamefont {Jeong}},\ and\
  \bibinfo {author} {\bibfnamefont {S.-J.}\ \bibnamefont {Kwon}},\ }\bibfield
  {title} {\bibinfo {title} {Orthogonal golay code based ultrasonic imaging
  without reducing frame rate},\ }in\ \href@noop {} {\emph {\bibinfo
  {booktitle} {2002 IEEE Ultrasonics Symposium, 2002. Proceedings.}}},\
  Vol.~\bibinfo {volume} {2}\ (\bibinfo {organization} {IEEE},\ \bibinfo {year}
  {2002})\ pp.\ \bibinfo {pages} {1705--1708}\BibitemShut {NoStop}%
\bibitem [{\citenamefont {Zeng}\ and\ \citenamefont
  {McGough}(2008)}]{Zeng2008}%
  \BibitemOpen
  \bibfield  {author} {\bibinfo {author} {\bibfnamefont {X.}~\bibnamefont
  {Zeng}}\ and\ \bibinfo {author} {\bibfnamefont {R.~J.}\ \bibnamefont
  {McGough}},\ }\bibfield  {title} {\bibinfo {title} {Evaluation of the angular
  spectrum approach for simulations of near-field pressures},\ }\href@noop {}
  {\bibfield  {journal} {\bibinfo  {journal} {The Journal of the Acoustical
  Society of America}\ }\textbf {\bibinfo {volume} {123}},\ \bibinfo {pages}
  {68} (\bibinfo {year} {2008})}\BibitemShut {NoStop}%
\bibitem [{\citenamefont {Fong}\ and\ \citenamefont
  {Saunders}(2011)}]{fong2011lsmr}%
  \BibitemOpen
  \bibfield  {author} {\bibinfo {author} {\bibfnamefont {D.~C.-L.}\
  \bibnamefont {Fong}}\ and\ \bibinfo {author} {\bibfnamefont {M.}~\bibnamefont
  {Saunders}},\ }\bibfield  {title} {\bibinfo {title} {Lsmr: An iterative
  algorithm for sparse least-squares problems},\ }\href@noop {} {\bibfield
  {journal} {\bibinfo  {journal} {SIAM Journal on Scientific Computing}\
  }\textbf {\bibinfo {volume} {33}},\ \bibinfo {pages} {2950} (\bibinfo {year}
  {2011})}\BibitemShut {NoStop}%
\bibitem [{\citenamefont {Paige}\ and\ \citenamefont
  {Saunders}(1982)}]{paige1982lsqr}%
  \BibitemOpen
  \bibfield  {author} {\bibinfo {author} {\bibfnamefont {C.~C.}\ \bibnamefont
  {Paige}}\ and\ \bibinfo {author} {\bibfnamefont {M.~A.}\ \bibnamefont
  {Saunders}},\ }\bibfield  {title} {\bibinfo {title} {Lsqr: An algorithm for
  sparse linear equations and sparse least squares},\ }\href@noop {} {\bibfield
   {journal} {\bibinfo  {journal} {ACM Transactions on Mathematical Software
  (TOMS)}\ }\textbf {\bibinfo {volume} {8}},\ \bibinfo {pages} {43} (\bibinfo
  {year} {1982})}\BibitemShut {NoStop}%
\bibitem [{\citenamefont {Berthon}\ \emph {et~al.}(2018)\citenamefont
  {Berthon}, \citenamefont {Morichau-Beauchant}, \citenamefont {Por{\'e}e},
  \citenamefont {Garofalakis}, \citenamefont {Tavitian}, \citenamefont
  {Tanter},\ and\ \citenamefont {Provost}}]{berthon2018spatiotemporal}%
  \BibitemOpen
  \bibfield  {author} {\bibinfo {author} {\bibfnamefont {B.}~\bibnamefont
  {Berthon}}, \bibinfo {author} {\bibfnamefont {P.}~\bibnamefont
  {Morichau-Beauchant}}, \bibinfo {author} {\bibfnamefont {J.}~\bibnamefont
  {Por{\'e}e}}, \bibinfo {author} {\bibfnamefont {A.}~\bibnamefont
  {Garofalakis}}, \bibinfo {author} {\bibfnamefont {B.}~\bibnamefont
  {Tavitian}}, \bibinfo {author} {\bibfnamefont {M.}~\bibnamefont {Tanter}},\
  and\ \bibinfo {author} {\bibfnamefont {J.}~\bibnamefont {Provost}},\
  }\bibfield  {title} {\bibinfo {title} {Spatiotemporal matrix image formation
  for programmable ultrasound scanners},\ }\href@noop {} {\bibfield  {journal}
  {\bibinfo  {journal} {Physics in Medicine \& Biology}\ }\textbf {\bibinfo
  {volume} {63}},\ \bibinfo {pages} {03NT03} (\bibinfo {year}
  {2018})}\BibitemShut {NoStop}%
\bibitem [{\citenamefont {Bioucas-Dias}\ and\ \citenamefont
  {Figueiredo}(2007)}]{bioucas2007new}%
  \BibitemOpen
  \bibfield  {author} {\bibinfo {author} {\bibfnamefont {J.~M.}\ \bibnamefont
  {Bioucas-Dias}}\ and\ \bibinfo {author} {\bibfnamefont {M.~A.}\ \bibnamefont
  {Figueiredo}},\ }\bibfield  {title} {\bibinfo {title} {A new twist: Two-step
  iterative shrinkage/thresholding algorithms for image restoration},\
  }\href@noop {} {\bibfield  {journal} {\bibinfo  {journal} {IEEE Transactions
  on Image processing}\ }\textbf {\bibinfo {volume} {16}},\ \bibinfo {pages}
  {2992} (\bibinfo {year} {2007})}\BibitemShut {NoStop}%
\bibitem [{\citenamefont {Figueiredo}\ \emph {et~al.}(2007)\citenamefont
  {Figueiredo}, \citenamefont {Nowak},\ and\ \citenamefont
  {Wright}}]{figueiredo2007gradient}%
  \BibitemOpen
  \bibfield  {author} {\bibinfo {author} {\bibfnamefont {M.~A.}\ \bibnamefont
  {Figueiredo}}, \bibinfo {author} {\bibfnamefont {R.~D.}\ \bibnamefont
  {Nowak}},\ and\ \bibinfo {author} {\bibfnamefont {S.~J.}\ \bibnamefont
  {Wright}},\ }\bibfield  {title} {\bibinfo {title} {Gradient projection for
  sparse reconstruction: Application to compressed sensing and other inverse
  problems},\ }\href@noop {} {\bibfield  {journal} {\bibinfo  {journal} {IEEE
  Journal of selected topics in signal processing}\ }\textbf {\bibinfo {volume}
  {1}},\ \bibinfo {pages} {586} (\bibinfo {year} {2007})}\BibitemShut {NoStop}%
\bibitem [{\citenamefont {Wear}(2022)}]{wear2022spatiotemporal}%
  \BibitemOpen
  \bibfield  {author} {\bibinfo {author} {\bibfnamefont {K.~A.}\ \bibnamefont
  {Wear}},\ }\bibfield  {title} {\bibinfo {title} {Spatiotemporal deconvolution
  of hydrophone response for linear and nonlinear beams—part i: Theory,
  spatial-averaging correction formulas, and criteria for sensitive element
  size},\ }\href@noop {} {\bibfield  {journal} {\bibinfo  {journal} {IEEE
  transactions on ultrasonics, ferroelectrics, and frequency control}\ }\textbf
  {\bibinfo {volume} {69}},\ \bibinfo {pages} {1243} (\bibinfo {year}
  {2022})}\BibitemShut {NoStop}%
\bibitem [{\citenamefont {Dogan}\ \emph {et~al.}(2021)\citenamefont {Dogan},
  \citenamefont {Kruizinga}, \citenamefont {Bosch},\ and\ \citenamefont
  {Leus}}]{dogan2021multiple}%
  \BibitemOpen
  \bibfield  {author} {\bibinfo {author} {\bibfnamefont {D.}~\bibnamefont
  {Dogan}}, \bibinfo {author} {\bibfnamefont {P.}~\bibnamefont {Kruizinga}},
  \bibinfo {author} {\bibfnamefont {J.~G.}\ \bibnamefont {Bosch}},\ and\
  \bibinfo {author} {\bibfnamefont {G.}~\bibnamefont {Leus}},\ }\bibfield
  {title} {\bibinfo {title} {Multiple measurement vector model for
  sparsity-based vascular ultrasound imaging},\ }in\ \href@noop {} {\emph
  {\bibinfo {booktitle} {2021 IEEE Statistical Signal Processing Workshop
  (SSP)}}}\ (\bibinfo {organization} {IEEE},\ \bibinfo {year} {2021})\ pp.\
  \bibinfo {pages} {501--505}\BibitemShut {NoStop}%
\bibitem [{\citenamefont {Bar-Zion}\ \emph {et~al.}(2018)\citenamefont
  {Bar-Zion}, \citenamefont {Solomon}, \citenamefont {Tremblay-Darveau},
  \citenamefont {Adam},\ and\ \citenamefont {Eldar}}]{bar2018sushi}%
  \BibitemOpen
  \bibfield  {author} {\bibinfo {author} {\bibfnamefont {A.}~\bibnamefont
  {Bar-Zion}}, \bibinfo {author} {\bibfnamefont {O.}~\bibnamefont {Solomon}},
  \bibinfo {author} {\bibfnamefont {C.}~\bibnamefont {Tremblay-Darveau}},
  \bibinfo {author} {\bibfnamefont {D.}~\bibnamefont {Adam}},\ and\ \bibinfo
  {author} {\bibfnamefont {Y.~C.}\ \bibnamefont {Eldar}},\ }\bibfield  {title}
  {\bibinfo {title} {Sushi: Sparsity-based ultrasound super-resolution
  hemodynamic imaging},\ }\href@noop {} {\bibfield  {journal} {\bibinfo
  {journal} {IEEE transactions on ultrasonics, ferroelectrics, and frequency
  control}\ }\textbf {\bibinfo {volume} {65}},\ \bibinfo {pages} {2365}
  (\bibinfo {year} {2018})}\BibitemShut {NoStop}%
\bibitem [{\citenamefont {van~der Meulen}\ \emph {et~al.}(2018)\citenamefont
  {van~der Meulen}, \citenamefont {Kruizinga}, \citenamefont {Bosch},\ and\
  \citenamefont {Leus}}]{van2018calibration}%
  \BibitemOpen
  \bibfield  {author} {\bibinfo {author} {\bibfnamefont {P.}~\bibnamefont
  {van~der Meulen}}, \bibinfo {author} {\bibfnamefont {P.}~\bibnamefont
  {Kruizinga}}, \bibinfo {author} {\bibfnamefont {J.~G.}\ \bibnamefont
  {Bosch}},\ and\ \bibinfo {author} {\bibfnamefont {G.}~\bibnamefont {Leus}},\
  }\bibfield  {title} {\bibinfo {title} {Calibration techniques for
  single-sensor ultrasound imaging with a coding mask},\ }in\ \href@noop {}
  {\emph {\bibinfo {booktitle} {2018 52nd Asilomar Conference on Signals,
  Systems, and Computers}}}\ (\bibinfo {organization} {IEEE},\ \bibinfo {year}
  {2018})\ pp.\ \bibinfo {pages} {1641--1645}\BibitemShut {NoStop}%
\end{thebibliography}%

\begin{figure*}
    \centering
    \includegraphics[width=2\columnwidth]{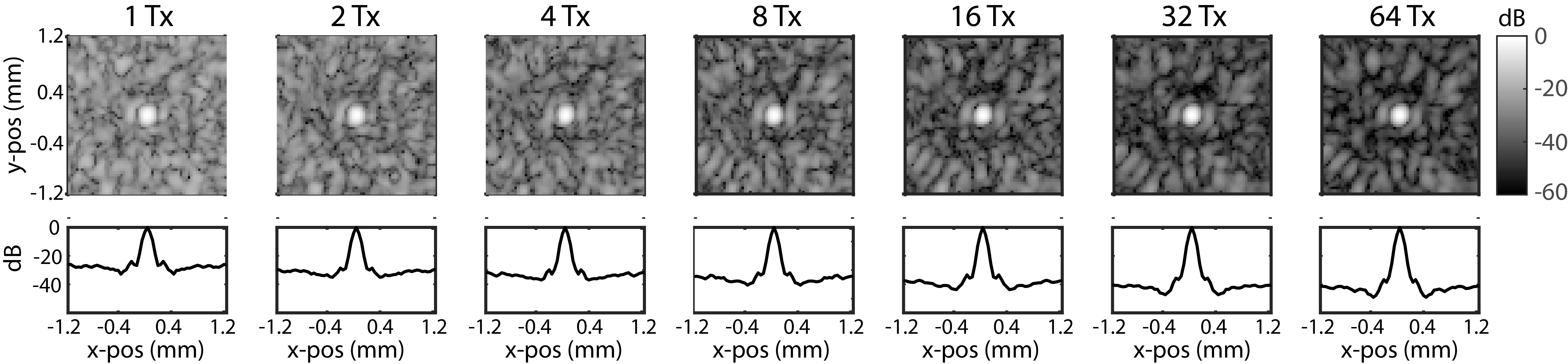}
    \caption{Lateral correlations for a voxel at a fixed (x,y,z) position for increasing transmission number. Compounding additional orthogonal transmissions decreases these correlations and improves the image reconstruction.}
    \label{SFig_1}
\end{figure*}

\begin{figure*}
    \centering
    \includegraphics[width=2\columnwidth]{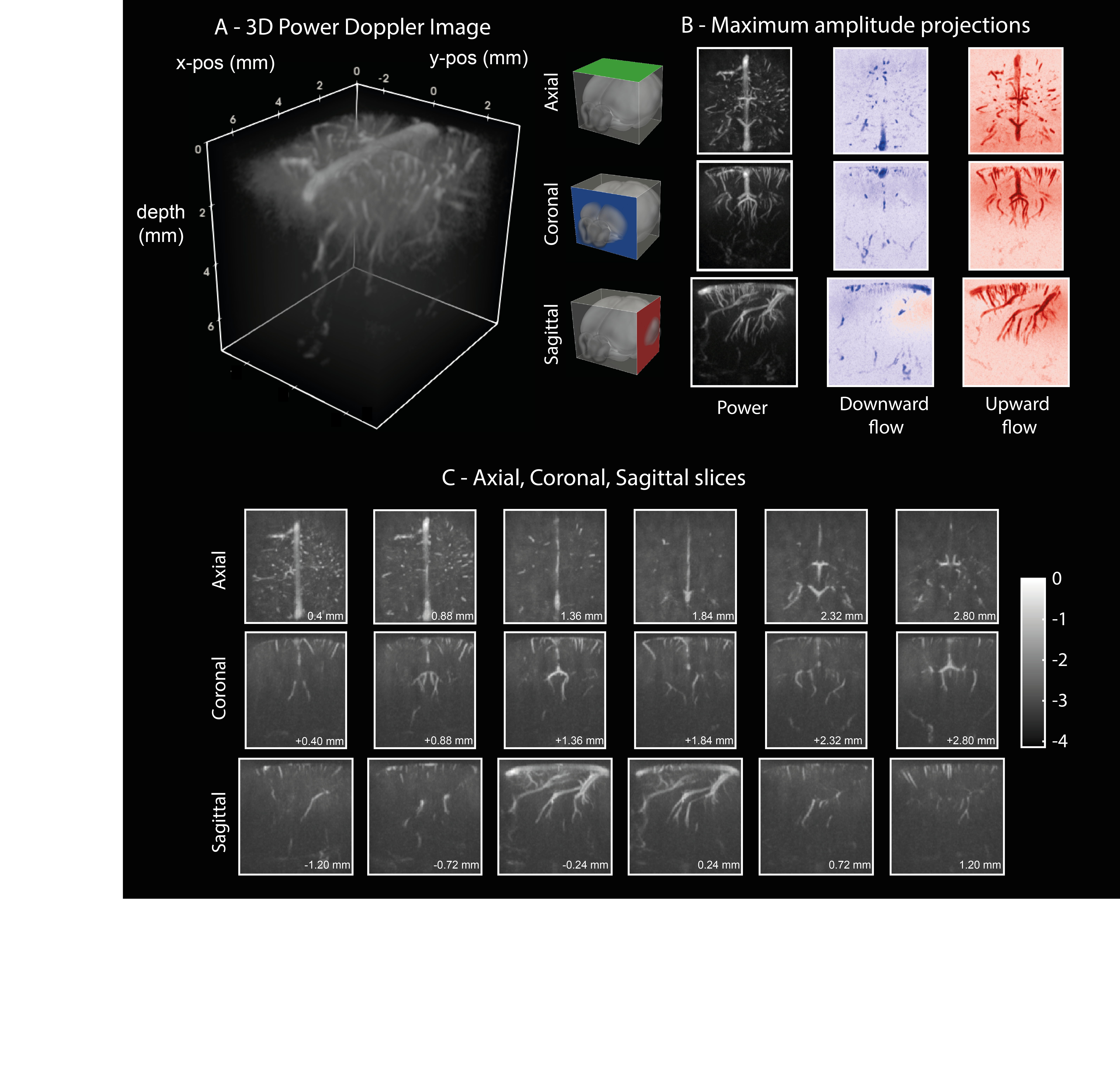}
    \caption{Results of haemodynamic imaging in the awake mouse brain. (A) 3D rendering of the reconstructed PDI of the awake mouse brain. Image was formed by compounding a filtered data-set comprised of 7679 volumes. (B) Axial, coronal, and sagittal maximum amplitude projections through the reconstructed Power (left) and Colour Doppler volumes. (C) Sub-projections through the PDI rendered in (A). Each slice was formed from a maximum amplitude projection through a set of planes with a thickness of 480 $\mu m$ (corresponding to 12 planes in the reconstructed volume). For the coronal slices the positions are indicated relative to Bregma. Compared to the anesthetised data shown in Fig. \ref{Figure_2} the awake data resolves a lower number of cortical vessels and less vessels deeper in the brain due to the attenuation generated by the TPX film. The field of view is also smaller due to the smaller cranio window. }
    \label{SFig_2}
\end{figure*}
\begin{figure*}
    \centering
    \includegraphics[width=2\columnwidth]{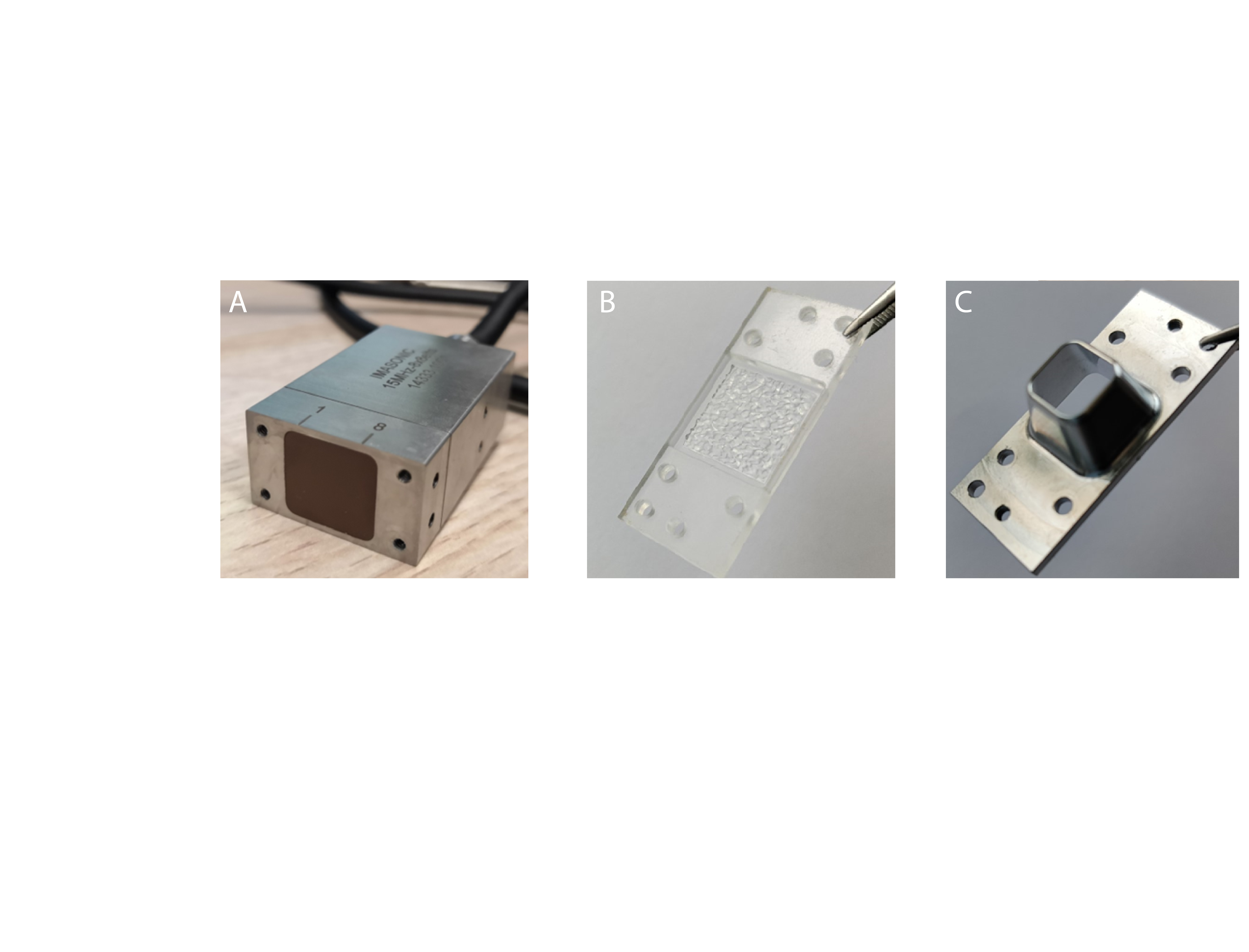}
    \caption{A: photo of 8$\times$8 matrix probe used for imaging experiments. B: photo of Rexolite spatial encoding mask fabricated using CNC micromachining. C: photograph of Aluminium waveguide fabricated using CNC micromachining. }
    \label{SFig_3}
\end{figure*}

\begin{figure*}
    \centering
    \includegraphics[width=2\columnwidth]{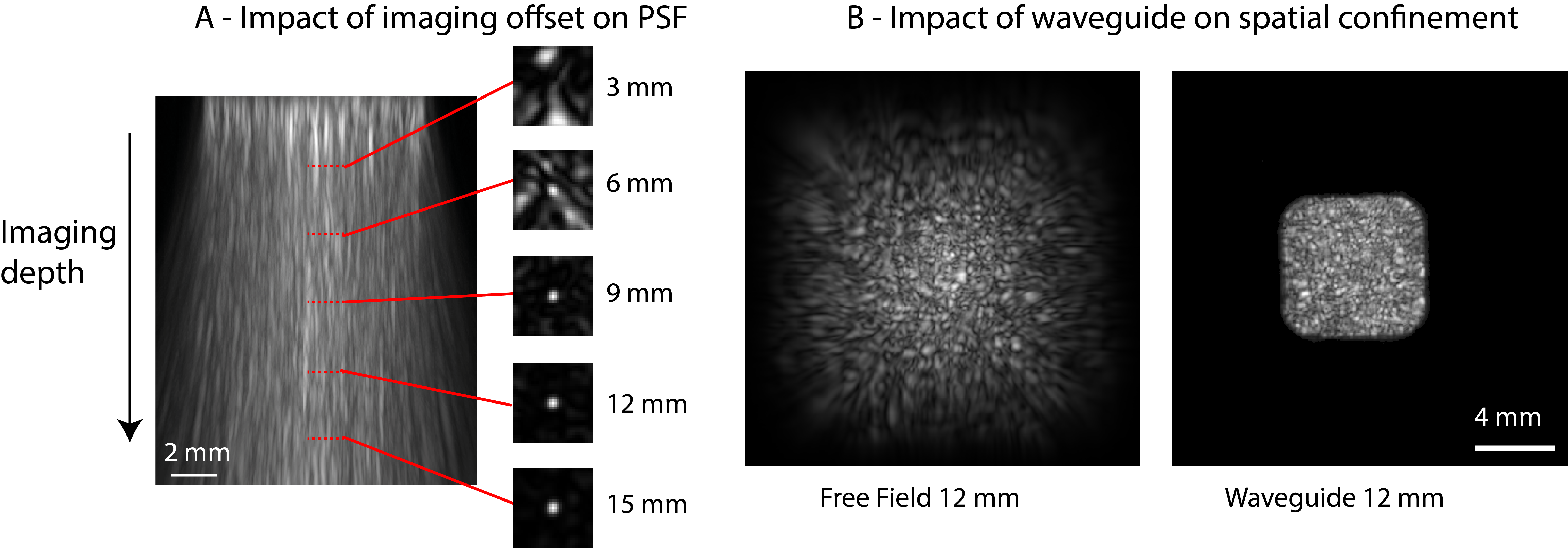}
    \caption{(A) Impact of imaging off-set on the point spread function. The top image is a maximum amplitude projection of a field at 15.6 MHz experimentally measured through an example coding mask while transmitting a plane wave. For this measurement no wave-guide was added to the coding mask so the field freely diverges. The 5 images on the bottom show the (normalised) correlations for a fixed (x,y) pixel at 5 different depths (3, 6, 9, 12, 15 mm) from the front surface of the probe. These correlations were analysed for a synthetic aperture transmission scheme as employed in the experimental measurements. For planes close to the imaging probe the lack of overlap between the fields of neighbouring elements makes it impossible to resolve echoes from distinct (x,y) locations. (B) Experimentally measured fields for a coding mask without (left) and with (right) a waveguide transmitting a plane-wave on the matrix probe. The waveguide effectively confines the transmitted field to the desired imaging aperture.}
    \label{SFig_4}
\end{figure*}

\begin{figure*}
    \centering
    \includegraphics[width=2\columnwidth]{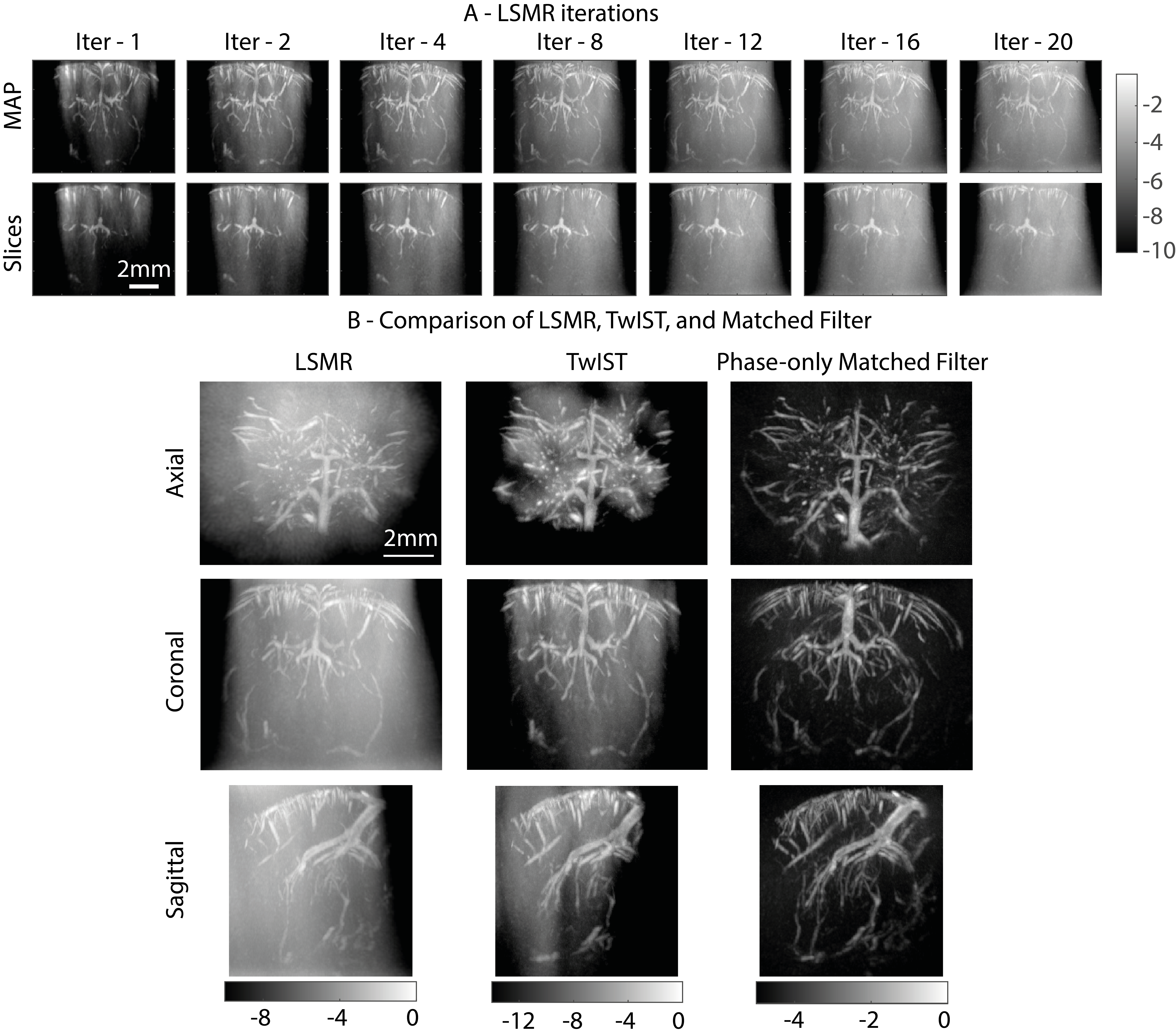}
    \caption{(A) Maximum amplitude projections (top row) and slices (bottom row) through PDIs reconstructed with different iteration numbers of the LSMR algorithm. All of the volumes formed in the figure were formed by reconstructing the full 8041 frames comprising the anesthetised data-set, however, the data was filtered differently to the images presented in Fig.\ \ref{Figure_2} and Fig.\ \ref{SFig_2}. Only the 5501-6500 spatial singular vectors were used to increase the computational efficiency of the iterative methods. The slices were formed from maximum amplitude projections through an 800 $\mu m$ coronal cross section. With increasing iteration number the spatial variation in the amplitude diminishes at the expense of diminished contrast. (B) Axial, coronal, and sagittal maximum amplitude projections of PDIs formed with the three different reconstruction methods evaluated in this work. The LSMR projection was taken from iteration 16, the TwIST algorithm was first run to convergence followed by the use of the conjugate gradient method to debias the final volume.}
    \label{SFig_5}
\end{figure*}

\begin{figure*}
    \centering
    \includegraphics[width=2\columnwidth]{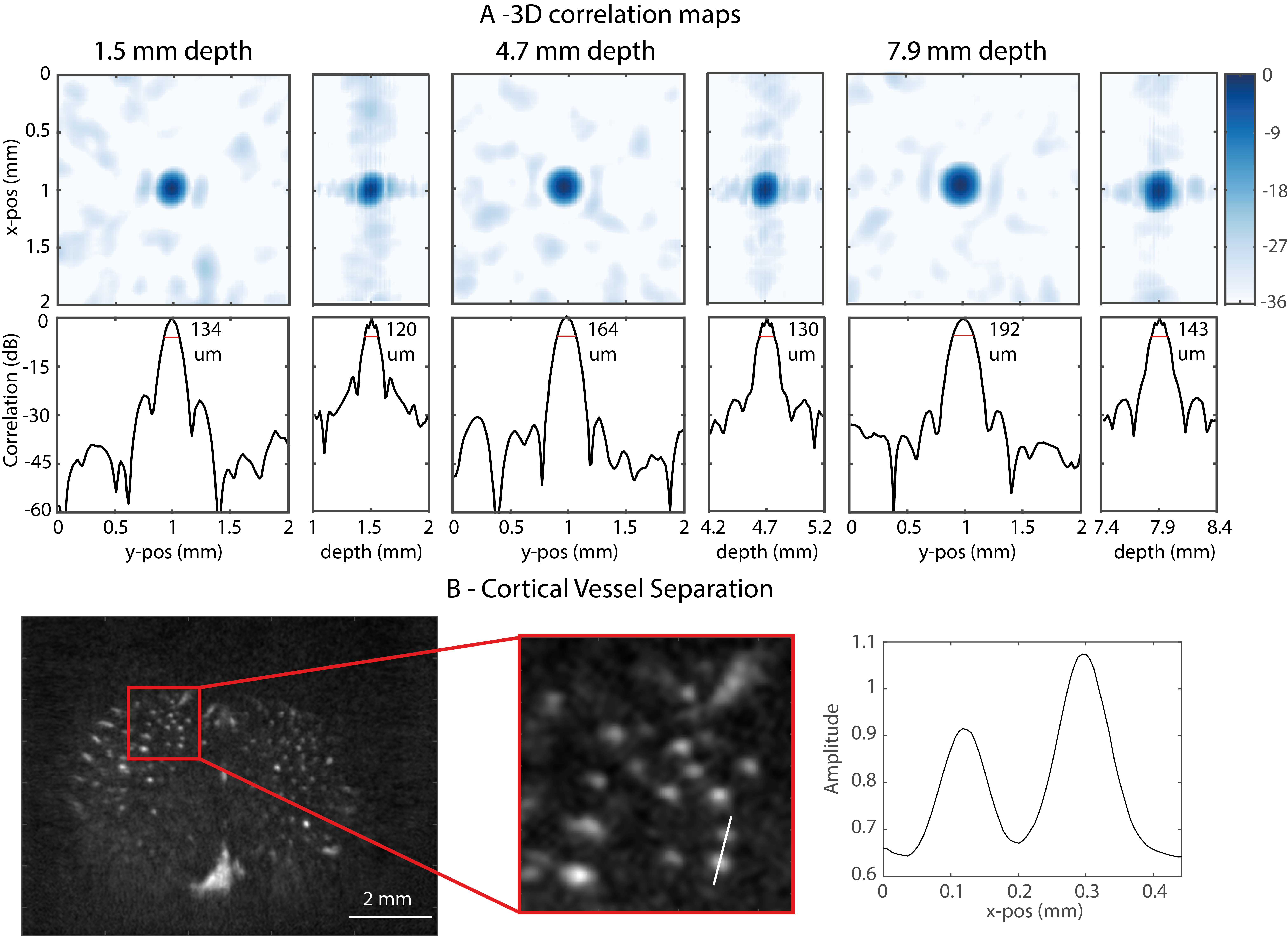}
    \caption{(A) Analysis of the axial and lateral correlations of the matrix $A$ with a fixed position $(x,y)$ at a 3 different depths. The correlations were evaluated on a grid with a 40$\mu m$ step size over a 2$\times$2$\times$1 mm volume centred on the voxel of interest. The top-row shows maximum amplitude projections through the spatial correlations over this volume at each depth. The bottom row plots lateral and axial cross-sections through the point of interest for each depth. The values in each sub-figure of the bottom row denote the -6dB width. (B) Assessment of vessel resolution in the cortex approximately 0.6 $mm$ inside the mouse brain. Separations of less than 200 $\mu m$ can be identified supporting the assessment of the spatial resolution from the system matrix. }
    \label{SFig_6}
\end{figure*}

\end{document}